\documentclass[aps,prb,reprint,superscriptaddress]{revtex4-1}
\usepackage{amsmath}
\usepackage{amsfonts}
\usepackage{graphicx}
\usepackage{hyperref}
\usepackage{color}
\newcommand{\real}[1] {\mathrm{Re}\, #1 \,}
\newcommand{\imag}[1]{\mathrm{Im}\, #1 \,}

\newcommand{\ud}{\mathrm{d}}

\newcommand{\ee}{\varepsilon}

\newcommand{\xpect}[1]{\left<  #1 \right>}

\newcommand{\rgs}[3]{G_{#1}^{#2}(#3)}
\newcommand{\gs}[3]{\mathcal{G}_{#1}^{#2}(#3)}
\newcommand{\gz}[3]{g_{#1}^{#2}(#3)}
\newcommand{\rgd}[1]{G_{d}(#1) }

\newcommand{\RGs}[2]{G_{#1}(#2) }

\newcommand{\Gz}[2]{g_{#1}(#2) }
\newcommand{\Seff}[2]{\Sigma_{\text{eff}}^{#1}(#2)}
\newcommand{\Teff}[2]{T_{\text{eff}}^{#1}(#2)}
\newcommand{\SE}[1]{\Sigma(#1)}
\newcommand{\trace}[1]{\mathrm{Tr}\big\{ #1 \big\}}
\newcommand{\TM}[1]{\mathrm{T}(#1)}
\newcommand{\Hg}{H_{\mathrm{G}}}
\newcommand{\Himp}{H_{\mathrm{I}}}
\newcommand{\HtopA}{H_{\text{T-G}}}
\newcommand{\HvacA}{H_{\text{V-G}}}
\newcommand{\HrecA}{H_{\text{R-G}}}
\newcommand{\Hhs}{H_{\text{H-G}}}
\newcommand{\Hig}{H_{\text{I-G}}}
\newcommand{\uu}{\hat{\mathbf{u}}}

\renewcommand{\vec}[1]{\boldsymbol{\mathbf{#1}}}
\renewcommand{\exp}[1]{\mathrm{e}^{#1}}

\begin{document}

\title{Symmetry--protected coherent transport for diluted vacancies and adatoms in graphene}

\author{David A.\ Ruiz-Tijerina}
\affiliation{Instituto de F\'{\i}sica, Universidade de S\~{a}o Paulo,
C.P.\ 66318, 05315--970 S\~{a}o Paulo, SP, Brazil}

\author{Luis G.\ G.\ V.\ Dias da Silva}
\affiliation{Instituto de F\'{\i}sica, Universidade de S\~{a}o Paulo,
C.P.\ 66318, 05315--970 S\~{a}o Paulo, SP, Brazil}

\date{\today}

\begin{abstract}
We study the effects of a low concentration of adatoms or single vacancies in the linear--response transport properties of otherwise clean graphene. These impurities were treated as localized orbitals, and for each type two cases with distinct coupling symmetries were studied. For adatoms, we considered top-- and hollow--site adsorbates (TOP and HS). For vacancies, we studied impurity formation by soft bond reconstruction (REC), as well as the more symmetric case of charge accumulation in unreconstructed vacancies (VAC).  Our results indicate that the transport is determined by usual impurity scattering when the graphene-impurity coupling does not possess $C_{3v}$ symmetry (TOP and REC). In contrast, VAC impurities decouple from the electronic states at the Dirac points, and yield no contribution to the resistivity for a sample in charge neutrality. Furthermore, the inversion--symmetry--conserving HS impurities also decouple from entire sets of momenta throughout the Brillouin zone, and do not contribute to the resistivity within a broad range of parameters. These behaviors are protected by $C_{3v}$ and inversion symmetry, respectively, and persist for more general impurity models.
\end{abstract}

\pacs{}
\maketitle

\section{Introduction}\label{sec:intro}

Graphene has been hailed as a promising material due to its unique electronic transport properties,\cite{castro-neto_rmp_2009} governed by elementary excitations that behave as  massless, chiral Dirac fermions.\cite{DasSarma:Rev.Mod.Phys.:407:2011} This leads to unique features, such as a universal minimum ballistic conductance\cite{Katsnelson:Eur.Phys.J.B:157:2006,Tworzydlo:Phys.Rev.Lett.:246802:2006,ferreira_prl_2015}  $\sigma_{\rm min}= 4e^2/\pi h$,  expected for an infinite ``clean" graphene sample at the charge--degeneracy point, and which has been confirmed by experiments.\cite{Heersche:Nature:56:2007,Miao:Science:1530:2007} 

The effects of short--range disorder (scattering centers, such as impurities or defects) in the transport properties of graphene has been the subject of intense research efforts. It has been argued that the conductivity depends strongly on the nature of the scattering processes, \cite{ostrovsky_prb_2006,ostrovsky_prl_2010,pachoud_prb_2014} and particularly on the symmetries of the disorder distribution.\cite{asmar_prb_2015} Moreover, it has been also established that impurities or defects that break the sublattice symmetry produce intervalley scattering and lead to non--universal conductivities. \cite{Stauber:Phys.Rev.B:205423:2007,hentschel_prb_2007}
In fact, experimental results for irradiated graphene show a strong decrease in the minimum conductivity, well below the universal value, indicating the onset of an insulating behavior.\cite{Chen:Phys.Rev.Lett.:236805:2009}

More recently, there has been rising interest in the formation of magnetic moments in impurities and vacancies in graphene,\cite{andrei_natphys_2016,Zhang2016,Gonzalez-Herrero2016} and the possible observation of the Kondo effect. \cite{Chen:NaturePhys.:535:2011,Miranda14} In this context, it has become clear that it is crucial considering not only the type of impurity (e.g., adatom or defect) but, more importantly, \textit{how} it couples to the graphene lattice.

For instance, Uchoa et al. \cite{uchoa_prl_2011} find that the effective impurity hybridization strongly depends on whether the graphene-impurity coupling breaks or preserves the $C_{3v}$ point group symmetry of the sublattice. If this symmetry is preserved, the hybridization function is strongly suppressed near the charge--degeneracy point ($\epsilon\!=\!0$), since it scales with energy as $|\epsilon|^3$. In contrast, $C_{3v}$--breaking couplings scale as\cite{uchoa_prl_2011} $|\epsilon|$. In fact, this effective decoupling of the impurities at low energies has been studied in the context of disorder in graphene, leading to so--called ``anomalous Anderson localization," in sharp contrast to symmetry--breaking impurities and Coulomb charge centers.\cite{Garcia:Phys.Rev.B:085425:2014}

In this paper we study the effects of a low impurity density in the transport properties of graphene, in terms of the symmetry of the impurity couplings. Using the Kubo formalism, we derive general expressions for the dc electrical resistivity in systems with different types of adatoms (top--site and hollow--site) and vacancies (symmetric and reconstructed), in varying concentrations.

The overall behaviors for impurities with different symmetry properties are strikingly different. While top--site adsorbates and reconstructed vacancies give a finite resistivity contribution, we find that hollow--site adatoms and symmetric vacancies decouple from the electronic states at the Dirac points, leading to a \emph{vanishing} contribution to the resistivity in charge--neutral graphene. Strikingly, hollow--site impurities also decouple from entire sets of states throughout the Brillouin zone, and do not contribute to the sample resistivity for any value of the carrier density. These are quantum--interference effects, and are protected by $C_{3v}$ and inversion symmetry, respectively. 

In systems with a mixture of symmetry--preserving and symmetry--breaking impurities (the most likely scenario in real experiments), we find that the impurity contribution to the resistivity can change by several orders of magnitude, depending on the relative symmetric/non--symmetric impurity concentration. Our results show a strong temperature dependence for the contribution of non--symmetric impurities. Since the graphene conductivity minimum (resistivity maximum) is temperature--independent down to 30 mK, it should be possible to detect such a robust impurity contribution in transport experiments. 

The remainder of this paper is organized as follows: Section \ref{sec:formalism} gives a brief introduction to graphene in the tight--binding approximation, and introduces the graphene-impurity couplings and their symmetries. In Section \ref{sec:linear_response} we develop a Kubo formalism for the low--energy regime, and derive a general formula for the resistivity at low impurity density. Numerical results for the resistivity of one-- and two--impurity mixtures as a function of temperature and chemical potential are presented and discussed in Sec\, \ref{sec:resistivity}. Finally, we present our conclusions in Sec.\ \ref{sec:conclusions}.

\section{Impurities in a graphene sample}\label{sec:formalism}
\begin{figure}[t]
\begin{center}
\includegraphics[width=0.95\columnwidth]{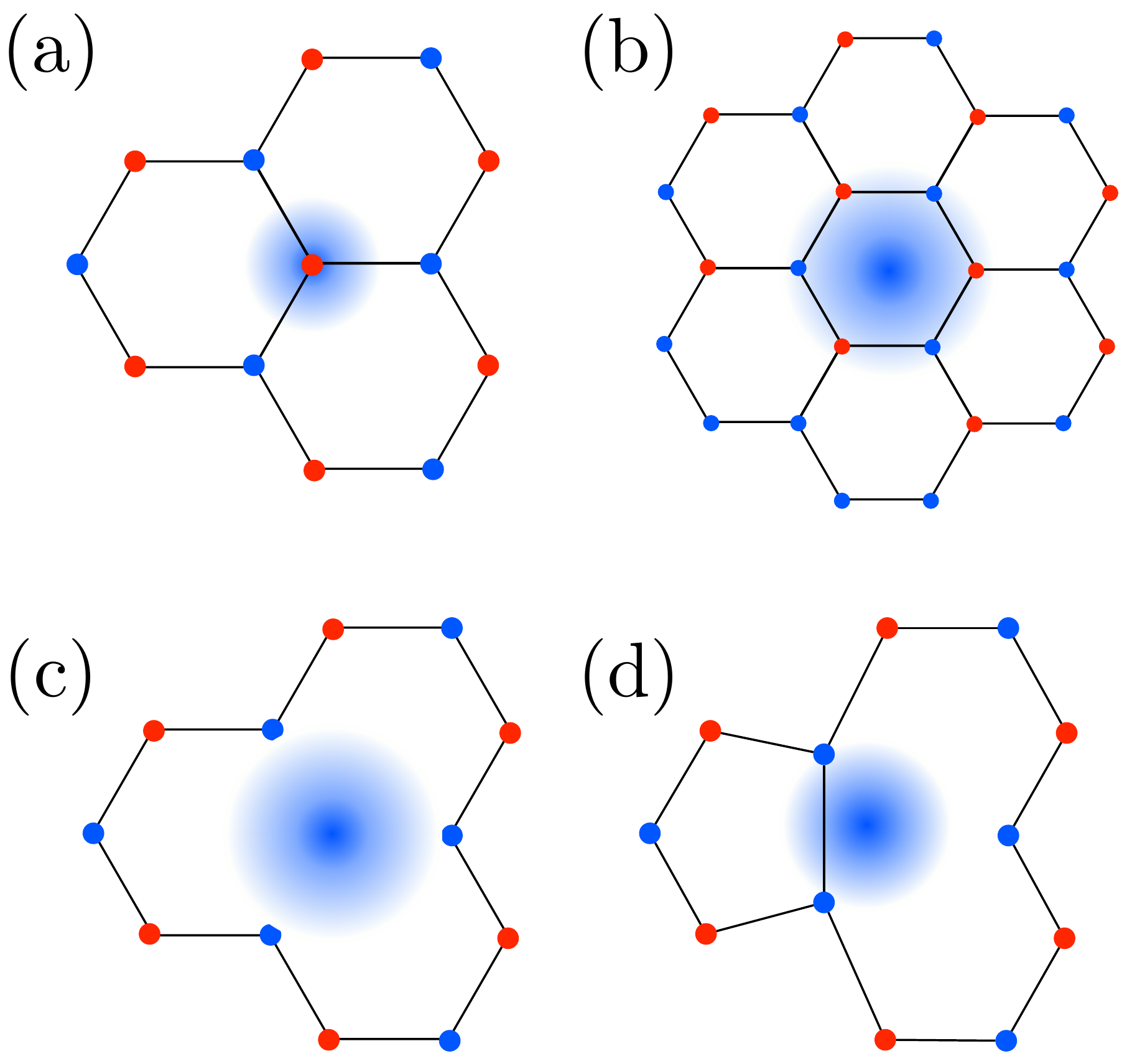}
\caption{(Color online) Single impurities in a graphene lattice: (a) Top--site adsorbate, (b) hollow--site $s$-level adsorbate, (c) symmetric vacancy and (d) reconstructed  vacancy. Lattice sites in red and blue belong to sublattices $A$ and $B$, respectively.}
\label{fig:impurities}
\end{center}
\end{figure}
Graphene can be described by the tight--binding Hamiltonian
\begin{equation}\label{eq:tight_binding}
    \Hg = -t\sum_{s}\sum_{\vec{R}_i}\sum_{j=1}^3\left\{ a_{s}^\dagger(\vec{R}_i) b_{s}(\vec{R}_i+a\uu_j) + \text{H.\ c.} \right\},
\end{equation}
where operators $a_s^\dagger(\vec{R}_i)$ [$a_s(\vec{R}_i)$] and $b_s^\dagger(\vec{R}_i)$ [$b_s(\vec{R}_i)$] create (annihilate) electrons of spin projection $s$ at the $i$th site of sublattices $A$ and $B$, respectively. The vector $\vec{R}_i$ in Eq.\ (\ref{eq:tight_binding}) and all expressions henceforth runs over sublattice $A$ sites. All nearest neighbors to these sites belong to sublattice $B$, and are located at positions $\vec{R}_i + a\uu_j$, where $\uu_1 = \hat{\mathbf{x}}$, $\uu_2 = -\hat{\mathbf{x}}/2 + \hat{\mathbf{y}}\sqrt{3}/2$ and $\uu_3 = -\hat{\mathbf{x}}/2 - \hat{\mathbf{y}}\sqrt{3}/2$ are unit vectors and $a$ is the nearest--neighbor spacing.

$\Hg$ can be expressed in terms of plane--wave operators as
\begin{equation}\label{eq:Hg_ab_k}
    \Hg = -t\sum_{\vec{k}s}\left\{ \Phi(\vec{k})\,a_{\vec{k},s}^\dagger b_{\vec{k},s} + \text{H.\ c.} \right\},
\end{equation}
with
\begin{equation}\label{eq:Phi}
    \Phi(\vec{k})\equiv\sum_{j=1}^3 \exp{ia\vec{k}\cdot\uu_j}.
\end{equation}
This model can be diagonalized exactly, giving two energy bands with dispersions $\ee_{\pm}(\vec{k}) = \pm t|\Phi(\vec{k})|$ and corresponding operators $c_{\pm,\vec{k}s}$. Defining the column vectors $c_{\vec{k},s}=(c_{+,\vec{k},s},\,c_{-,\vec{k},s})^T$ and $\psi_{\vec{k},s}=(a_{\vec{k},s},\,b_{\vec{k},s})^T$, the operators of the basis in which $\Hg$ is diagonal (hereafter referred to as the ``$c$-basis") can be related to those of the $A$ and $B$ sublattices through the unitary transformation $\psi_{\vec{k},s} = U_{\vec{k}}c_{\vec{k},s}$, with
\begin{equation}\label{eq:unitary_2band}
    U_{\vec{k}} = \frac{1}{\sqrt{2}}\begin{pmatrix}
    \frac{\Phi(\vec{k})}{|\Phi(\vec{k})|} & -\frac{\Phi(\vec{k})}{|\Phi(\vec{k})|}\\
    1 & 1\\
    \end{pmatrix}.
\end{equation}

A single (non--interacting) impurity will be treated as a localized level described by the term
\begin{equation}\label{eq:Himp}
    \Himp = \sum_{s}\ee_d\,d_{s}^\dagger d_{s},
\end{equation}
where $d_{s}^\dagger$ ($d_{s}$) creates (annihilates) an electron of spin projection $s$ in the impurity orbital. How this orbital couples to the graphene will depend on its geometry and its position in the lattice. In this article we consider two representative types of impurities: vacancies and adsorbates. Within these types, we will also divide the impurities by the local symmetry of their couplings to the graphene lattice. The impurity-graphene couplings in real space are described in Appendix \ref{app:impurities}. In the $c$-basis, these terms take the form\cite{uchoa_prl_2009,uchoa_prl_2011}
\begin{equation}
    \Hig = \sum_{s,\alpha}\sum_{\vec{k}}\left\{\Theta_{I}^\alpha(\vec{k})\,d^\dagger_s c_{\alpha,\vec{k},s} + \text{H.\ c.} \right\},
\end{equation}
where $\alpha=+,-$ indicates the band index and $I$  the impurity type.

\begin{figure}[t]
\begin{center}
\includegraphics[width=\columnwidth]{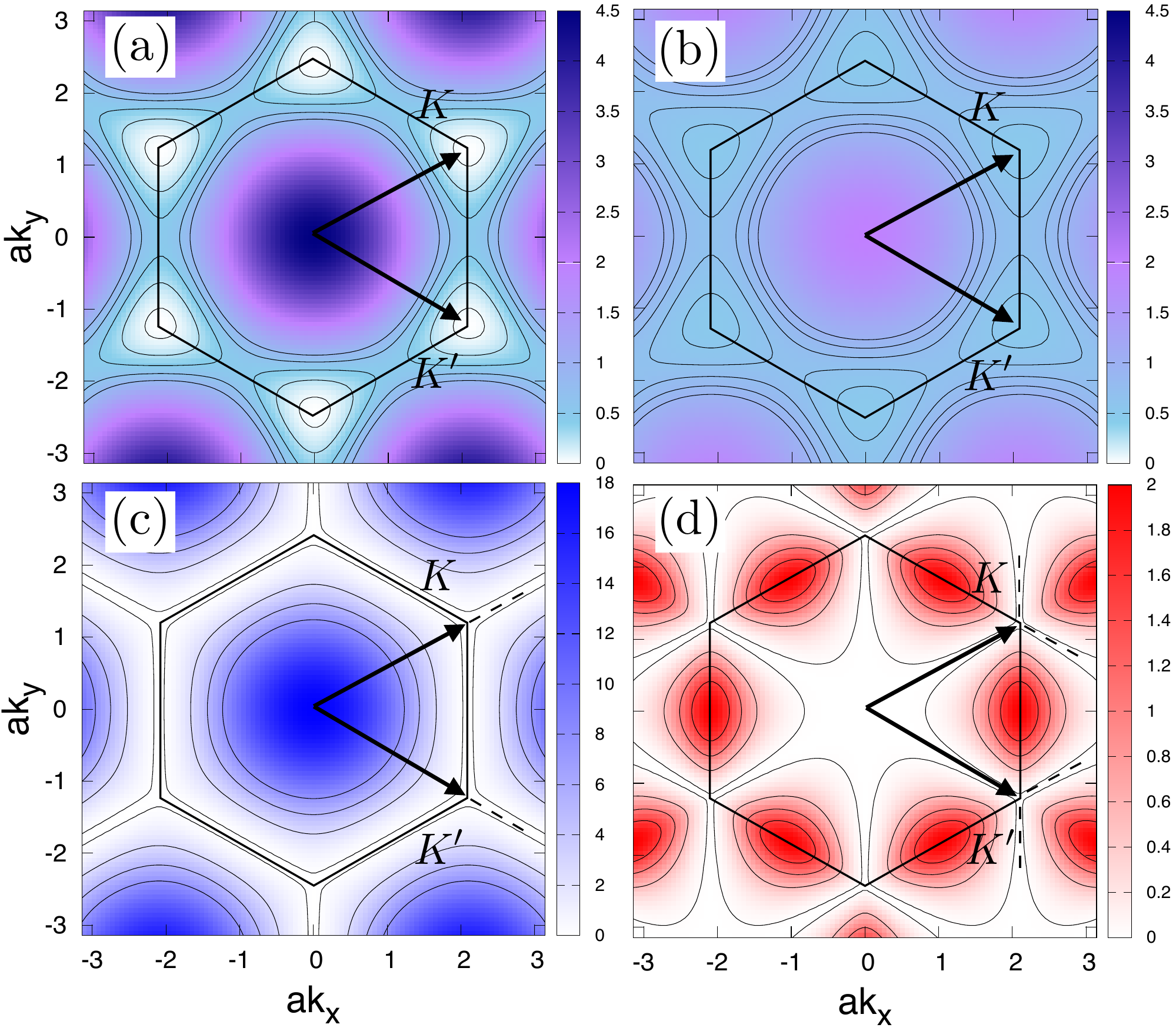}
\caption{(Color online) $|\Theta_I(\vec{k})|^2$ for (a) vacancies ($I=V$) and (b) reconstructed vacancies ($I=R$); (c) hollow--site adatom ($I=H$) upper band, and (d) hollow--site adatom lower band. The equipotential contours show topographic details at low energies, and the hexagons indicate the boundaries of the first Brillouin zone. The couplings vanish at $K$ and $K'$ for all cases except $I=R$, and only $|\Theta_{H}^\pm(\vec{k})|$ display line nodes at specific angles about these points.
}
\label{fig:thetasq}
\end{center}
\end{figure}

Vacancies are atomic--scale defects consisting of missing atoms in the graphene lattice.\cite{hashimoto_nature_2004,meyer_nanolett_2008,robertson_acsnano_2013,robertson_nanoscale_2013,andrei_natphys_2016} Charged vacancies (VAC) without bond reconstruction are the simplest from a geometrical point of view. When this kind of vacancy occurs in sublattice $A$, say, it will couple identically to all three surrounding carbons of sublattice $B$,\cite{ducastelle_prb_2013} as depicted in Fig.\ \ref{fig:impurities}(c). In a sense, sublattice $B$ is ``singled out'' and inversion symmetry is locally broken. Nonetheless, the point symmetry of this configuration matches that of the sublattice, and is encoded in the VAC-graphene momentum--space coupling
\begin{equation}\label{eq:vVAC}
    \Theta_{V}^\pm(\vec{k}) = \frac{V}{\sqrt{2}}\Phi(\vec{k}),
\end{equation}
which has $C_{3v}$ symmetry about the $K$ and $K'$ points [Fig.\ \ref{fig:thetasq}(a)]. Notice in particular that the coupling vanishes at these high--symmetry points. This is as a result of quantum interference, and is protected by $C_{3v}$ symmetry.

The situation is substantially different for vacancies with bond reconstruction (REC),\cite{barbary_prb_2003,ma_newjphys_2004,skowron_chemsocrev_2015} which in addition to not preserving local inversion also break the sublattice $C_{3v}$ point symmetry, as shown in Fig.\ \ref{fig:impurities}(d). The REC-graphene coupling is given by
\begin{equation}\label{eq:vREC}
    \Theta_{R,j}^\pm(\vec{k}) = \frac{V}{\sqrt{2}}\exp{ia\vec{k}\cdot\hat{\vec{u}}_j}\sum_{l\ne j}\exp{-ia\vec{k}\cdot\hat{\vec{u}}_l},
\end{equation}
which is explicitly not invariant under $C_3$ rotations. In deriving Eq.\ (\ref{eq:vREC}) we have placed the impurity orbital specifically at $a\hat{\vec{u}}_j$ from the vacancy site. This represents the coupling between the $sp^2$ orbital of the atom at $a\hat{\vec{u}}_j$ and the $\pi$ orbitals of the atoms at $a\hat{\vec{u}}_i$ ($i \ne j$) discussed in Refs.\ [\onlinecite{barbary_prb_2003, kanao_jpsj_2012}]. This is only one of three possible configurations that occur with equal probability throughout the sample, and thus have to be averaged to properly describe an impurity distribution. The average
\begin{equation}\label{eq:recav}
\begin{split}
    |\Theta_R(\vec{k})|^2 &= \frac{1}{3}\sum_{j=1}^3|\Theta_{R,j}(\vec{k})|^2\\
    &= V^2\left[ 1+|\Phi(\vec{k})|^2 - \frac{2}{3}\real{\Phi^2(\vec{k})}\right],
\end{split}
\end{equation}
shown in Fig.\ \ref{fig:thetasq}(b) demonstrates that global $C_{3v}$ symmetry is recovered for an ensemble of REC impurities. However, in this case the coupling is finite for all momenta, and particularly at the $K$ and $K'$ points. The interference that gave rise to the zeros in the VAC case is removed by the symmetry breaking, and the $C_{3v}$ symmetry of the REC impurity distribution is only recovered in average. It is worthwhile mentioning that, although the details of the coupling Eq.\ (\ref{eq:vREC}) depend on the precise location of the impurity orbital within the vacancy, the average Eq.\ (\ref{eq:recav}) does not.

Top--adsorbed impurities\cite{eelbo_prl_2013,junkermeier_jphyschem_2013,lin_nanolett_2015} (TOP) are adatoms located outside the graphene plane, forming covalent bonds with a single carbon atom in the honeycomb lattice [Fig.\ \ref{fig:impurities}(a)]. By coupling only to one site, TOP impurities locally break inversion symmetry,\cite{cheianov_epl_2010} while  still preserving full rotational symmetry in the plane. Because we treat the TOP impurity as a point--like object, it will couple equally with all momentum states in the graphene sample as
\begin{equation}\label{eq:vTOP}
    \Theta_{T}^\pm(\vec{k}) = \frac{V}{\sqrt{2}}.
\end{equation}
Thus, although the impurity isotropy preserves the sublattice point symmetry, it is clear that this symmetry is not inherited by the TOP-graphene coupling function.

In contrast, a hollow--site adsorbate (HS) with an $s$ or $d_{z^2}$ valence orbital preserves the sublattice point symmetry. By coupling to both sublattices with equal strength, HS impurities are also invariant under inversion [Fig.\ \ref{fig:impurities}(b)]. The HS-graphene coupling function has the form
\begin{equation}\label{eq:vHS}
    \Theta_{H}^\pm(\vec{k}) = \frac{V}{\sqrt{2}}\left[ \Phi^*(\vec{k}) \pm \frac{\Phi^2(\vec{k})}{|\Phi(\vec{k})|} \right],
\end{equation}
which possesses full $C_{3v}$ symmetry about the high--symmetry points. This is shown in Figs.\ \ref{fig:thetasq}(d)  and \ref{fig:thetasq}(c). Strikingly, the coupling to the upper (lower) band vanishes for all $\vec{k}$ about the $K$ ($K'$) point at angles $\phi = -\pi/2,\,\pi/6,\, 5\pi/6$, as well as for $\phi = -5\pi/6,\,-\pi/6,\,\pi/2$ about the $K'$ ($K$) point. These nodes are produced by quantum interference involving momentum states from both sublattices, and in fact appear quite independently of the specific point symmetry of the impurity (Appendix \ref{app:inversion}).

Indeed, this will be a central point in our discussion of transport: The presence of nodes in the coupling function $\Theta_H(\vec{k})$ is protected by inversion symmetry. Even if the point symmetry of the coupling were reduced, inversion symmetry guarantees the presence of at least one node originating at each symmetry point. As we will show in Section \ref{sec:linear_response}, this decoupling of the impurity to graphene states throughout the Brillouin zone makes the impurity ``invisible'' to the overall transport behavior of the sample, resulting in a zero contribution to the resistivity.

In the following section we develop a Kubo formula for the resistivity of a graphene sample with an impurity density $n_{\text{imp}}\equiv N_{\text{imp}}/S$, where $N_{\text{imp}}$ is the number of impurities per unit cell and $S = 3a^2\sqrt{3}/2$ the (hexagonal) unit cell area. We will work in the dilute regime of $N_{\text{imp}} \ll 1$. The impurity type and symmetry will enter our formalism through the coupling functions introduced above.

\section{Linear--response transport}\label{sec:linear_response}
We are interested in exploring the regime of low impurity density of a mesoscopic graphene sample by means of standard transport measurements. At low temperatures and close to charge neutrality, the behavior of this system is determined entirely by momentum states near the $K$ and $K'$ points.

About these high--symmetry points the dispersion $\ee_\pm(\vec{K}^{(')}+\vec{k})=\pm t|\Phi(\vec{K}^{(')}+\vec{k})| \approx 3tak/2$ is linear and isotropic in $\vec{k}$. Furthermore, in the low--energy regime the valley index ($K$ or $K'$) behaves as an additional spin quantum number,\cite{castro-neto_rmp_2009} resulting in a model that describes Dirac quasiparticles. The vectors $\psi_{\vec{k}\sigma}$ and $c_{\vec{k}\sigma}$ become the 4-spinors\cite{ryu_prb_2009}
\begin{equation}
    {\small
    \psi_{\vec{k},s} = \begin{pmatrix}
    a_{\vec{K}+\vec{k},s} \\ b_{\vec{K}+\vec{k},s} \\ b_{\vec{K}'+\vec{k},s} \\ a_{\vec{K}'+\vec{k},s} 
    \end{pmatrix},\qquad
        c_{\vec{k},s} = \begin{pmatrix}
    c_{+,\vec{K}+\vec{k},s} \\ c_{-,\vec{K}+\vec{k},s} \\ c_{-,\vec{K}'+\vec{k},s} \\ c_{+,\vec{K}'+\vec{k},s}
    \end{pmatrix},
    }
\end{equation}
connected by the unitary transformation
\begin{equation}\label{eq:unitary}
    {\small
    U_{\vec{k}} = \frac{1}{\sqrt{2}}\begin{pmatrix}
    \frac{\Phi(\vec{K}+\vec{k})}{|\Phi(\vec{K}+\vec{k})|} & -\frac{\Phi(\vec{K}+\vec{k})}{|\Phi(\vec{K}+\vec{k})|} & 0 & 0\\
    1 & 1 & 0 & 0\\
    0 & 0 & 1 & 1\\
    0 & 0 &-\frac{\Phi(\vec{K}'+\vec{k})}{|\Phi(\vec{K}'+\vec{k})|} & \frac{\Phi(\vec{K}'+\vec{k})}{|\Phi(\vec{K}'+\vec{k})|},
    \end{pmatrix}.
    }
\end{equation}

\subsection{The current operator}\label{sec:num_op}
The current operator for momentum $\vec{q}$ is given by\cite{ostrovsky_prb_2006}
\begin{equation}\label{eq:current_abkkp}
    \vec{j}(\vec{q}) = ev_F\sum_{\vec{k},s}\psi_{\vec{k},s}^\dagger \tau^3\,\vec{\sigma}\,\psi_{\vec{k}+\vec{q},s},
\end{equation}
where $e$ is the electronic charge, $v_F = 3ta/2$ is the Fermi velocity ($\hbar = 1$) and $\tau^i$ ($\sigma^i$) are Pauli matrices acting on the valley (sublattice) subspace. In the $c$-basis the current components ($i=1,\,2$) take the form
\begin{equation}\label{eq:current_final}
    j^i(\vec{q}) = ev_F\sum_{\vec{k},s}c_{\vec{k},s}^\dagger\gamma^i(\vec{k},\vec{q})c_{\vec{k}+\vec{q},s},
\end{equation}
where $\gamma^i(\vec{k},\vec{q}) = U_{\vec{k}}^\dagger \tau^3 \, \sigma^i U_{\vec{k}+\vec{q}}$. For the remainder of this article we will work exclusively in this basis and omit the spin index $s$.  For zero momentum transfer ($\vec{q}\!=\!0$), the $\gamma^{i}$ matrices take the form:
\begin{widetext}
\begin{equation}
    \gamma^i(\vec{k},\vec{q}\!=\!0) = \begin{pmatrix}
\frac{\real{[\sigma^i_{12}\Phi^*(\vec{K}+\vec{k})]}}{|\Phi(\vec{K}+\vec{k})|} & \frac{i\imag{[\sigma^i_{12}\Phi^*(\vec{K}+\vec{k})]}}{|\Phi(\vec{K}+\vec{k})|} & 0 & 0\\
    -\frac{i\imag{[\sigma^i_{12}\Phi^*(\vec{K}+\vec{k})]}}{|\Phi(\vec{K}+\vec{k})|} & -\frac{\real{[\sigma^i_{12}\Phi^*(\vec{K}+\vec{k})]}}{|\Phi(\vec{K}+\vec{k})|} & 0 & 0\\
    0 & 0 & \frac{\real{[\sigma^i_{12}\Phi(\vec{K}'+\vec{k})]}}{|\Phi(\vec{K}'+\vec{k})|} & -\frac{i\imag{[\sigma^i_{12}\Phi(\vec{K}'+\vec{k})]}}{|\Phi(\vec{K}'+\vec{k})|}\\
    0 & 0 & \frac{i\imag{[\sigma^i_{12}\Phi(\vec{K}'+\vec{k})]}}{|\Phi(\vec{K}'+\vec{k})|} & -\frac{\real{[\sigma^i_{12}\Phi(\vec{K}'+\vec{k})]}}{|\Phi(\vec{K}'+\vec{k})|}
    \end{pmatrix}.
\end{equation}

\subsection{Kubo formula}\label{sec:sub_resistivity}
From Eq.\ (\ref{eq:current_final}) we can calculate the two--point current correlation function, and obtain the conductivity tensor in linear response via the Kubo formalism (Appendix \ref{app:kubo}). The resistivity tensor $\rho$ is then given by the inverse of the conductivity tensor as 
\begin{equation}\label{eq:resistivity}
    [\rho^{-1}]^{ij}(T)=\frac{(ev_F)^2}{\pi}\sum_{\vec{k}}\int\ud \omega\left[-\frac{\partial n_F(\omega,T,\mu)}{\partial \omega} \right]  
\trace{\gamma^i(\vec{k})\RGs{\vec{k}\vec{k}}{\omega^-}\gamma^j(\vec{k})\RGs{\vec{k}\vec{k}}{\omega^+}},
\end{equation}
\end{widetext}
where $n_F(\omega,T,\mu)$ is the Fermi-Dirac distribution for energy $\omega$, temperature $T$ and chemical potential $\mu$; $\RGs{\vec{k}\vec{k}}{\omega^+}$ [$\RGs{\vec{k}\vec{k}}{\omega^-}$] is the full retarded (advanced) graphene Green's function; and $\gamma^i(\vec{k}) = \gamma^i(\vec{k},\vec{q}=0)$.

Evaluating the trace in Eq.\ (\ref{eq:resistivity}) requires taking the matrix product for each impurity type. This process can be expedited by noting that the ratio $\Phi(\vec{K}^{(')}+\vec{k})/|\Phi(\vec{K}^{(')}+\vec{k})|$ in (\ref{eq:unitary}) is a function only of the momentum azimuthal angle $\phi$, to first order in $\vec{k}$. Therefore, the trace has the general form
\begin{equation}\label{eq:trace_general}
\begin{split}
    &A_{\mu\mu}^{ij}(\phi)\rgs{\vec{k}\vec{k}}{\mu\mu}{\omega^-}\rgs{\vec{k}\vec{k}}{\mu\mu}{\omega^+} + B_{\mu\nu}^{ij}(\phi)\rgs{\vec{k}\vec{k}}{\mu\mu}{\omega^-}\rgs{\vec{k}\vec{k}}{\nu\nu}{\omega^+}\\ +& C^{ij}_{\mu\nu}(\phi)\rgs{\vec{k}\vec{k}}{\mu\nu}{\omega^-}\rgs{\vec{k}\vec{k}}{\mu\nu}{\omega^+}+D^{ij}_{\mu\nu}(\phi)\rgs{\vec{k}\vec{k}}{\mu\nu}{\omega^-}\rgs{\vec{k}\vec{k}}{\nu\mu}{\omega^+},
\end{split}
\end{equation}
where the matrices $A^{ij},\,B^{ij},\,C^{ij},\,D^{ij}$ are impurity--dependent, and sums over repeated Greek indices are implied. This expression can be simplified by a few general considerations. First, if our result is to be valid for any uniform distribution of impurities in the dilute limit, the Green's functions in Eq.\ (\ref{eq:resistivity}) must be interpreted as the average over all possible uniform distributions. Assuming a very low impurity density, the self energy associated to the Green's function can be approximated as\cite{mahan}
\begin{equation}\label{eq:self_energy}
    \SE{\omega^\pm} = n_{\text{imp}}\TM{\omega^\pm}+\mathcal{O}( n_{\text{imp}}^2 ),
\end{equation}
where $\TM{\omega^\pm}$ is the single--impurity $T$ matrix. The $T$ matrix can be put in terms of the impurity local Green's function $\rgd{\omega^\pm}$ using the equation--of--motion method:\cite{zubarev,hewson_kpthf} 
\begin{equation}\label{eq:t_matrix}
    T_{\vec{k}\vec{k}'}^{\mu\nu}(\omega^\pm) = \Theta_I^\mu(\vec{k})\rgd{\omega^\pm}\Theta_I^\nu{}^*(\vec{k}'),
\end{equation}
where $\Theta_I^\mu$ are the elements of the $1\times4$ coupling matrix: $\Theta_I^1(\vec{k}) = \Theta_I^+(\vec{K}+\vec{k})$, $\Theta_I^2(\vec{k}) = \Theta_I^-(\vec{K}+\vec{k})$, $\Theta_I^3(\vec{k}) = \Theta_I^-(\vec{K}'+\vec{k})$ and $\Theta_I^4(\vec{k}) = \Theta_I^+(\vec{K}'+\vec{k})$. Using the symmetry properties of the coupling functions about $K$ and $K'$, and considering that the trace will be integrated over $\phi$, the sums in expression (\ref{eq:trace_general}) can be limited to  $\mu,\nu = 1,2$. Further, it can be shown that the intra--band--intra--valley terms $\rgs{\vec{k}\vec{k}}{\mu\mu}{\omega^-}\rgs{\vec{k}\vec{k}}{\mu\mu}{\omega^+}$ dominate in the dilute limit (Appendix \ref{app:Seff}), and we need to calculate only the prefactors $A^{ij}_{11}(\phi)$ and $A^{ij}_{22}(\phi)$.

With these approximations, the impurity contribution to the resistivity is simplified to
\begin{widetext}
\begin{equation}\label{eq:integral}
    [\rho^{-1}]^{ij} = \frac{2(ev_F)^2S}{(2\pi)^3}\int\ud^2k\int_{-\infty}^{\infty} \ud \omega \left[-\frac{\partial n_F(\omega,T,\mu)}{\partial \omega} \right]  \left[\frac{A^{ij}_{11}(\phi)\delta(\omega - v_Fk)}{n_{\text{imp}}|\Theta_I^1(\vec{k})|^2\rho_{dI}(\omega)}+\frac{A^{ij}_{22}(\phi)\delta(\omega + v_Fk)}{n_{\text{imp}}|\Theta_I^2(\vec{k})|^2\rho_{dI}(\omega)} \right],
\end{equation}
\end{widetext}
where $\rho_{dI}(\omega)$ is the spectral density for impurity type $I$.

For $I = T,\,R$ these integrals are always well behaved, but for $I = V$ the integrand has a singularity at each symmetry point [see Fig.\ \ref{fig:thetasq}(a)]. Most strikingly, for $I=H$ the integrand is singular at the symmetry points as well as at the line nodes shown in Figs.\ \ref{fig:thetasq}(c) and \ref{fig:thetasq}(d). These singularities lead to a diverging integral Eq.\ (\ref{eq:integral}).

This is one of the main results of this paper: Due to the $C_{3v}$ and inversion symmetries of their couplings, VAC and HS impurities are ``invisible'' to specific graphene momenta, which remain available for coherent transport and result in a zero impurity contribution to the system resistivity. As we will see below, a finite resistivity contribution due to impurity scattering is recovered away from charge neutrality for VAC impurities. This is not the case, however, for HS impurities, whose symmetry properties under inversion guarantee that states will be available for coherent transport at all energies.

We emphasize that Eq.\ (\ref{eq:integral}) represents the impurity contribution to the graphene resistivity, and its vanishing for VAC and HS impurities does not entail perfect conduction through the sample.\cite{mahan}  That said, our formalism could be paired with existing techniques that consider full--counting statistics, to determine the system transport under those conditions.\cite{titov_prl_2010,ostrovsky_prl_2010,Gattenloehner:Phys.Rev.Lett.:026802:2014} In addition, we remind the reader that these results are not valid for a high impurity concentration, where a determinant of whether the graphene symmetries are preserved is the symmetry of the impurity distribution itself.\cite{ostrovsky_prb_2006,asmar_prb_2015}

From an experimental point of view, when adsorbates are evaporated onto the graphene sheet both TOP and HS adatoms will be present in the sample.\cite{eelbo_prl_2013} A similar argument can be made for vacancies, where both VAC and REC sites will be created by, e.g., sputtering from an incident electron beam.\cite{robertson_natcomm_2012} With this in mind, our approach is to evaluate the average contribution to the longitudinal resistivity $\bar{\rho} \equiv [([\rho^{-1}]^{11} + [\rho^{-1}]^{22})/2]^{-1}$ of a graphene sample with two kinds of adatoms (TOP and HS), or two kinds of vacancies (REC and VAC). 

Given their unique symmetry properties, we will focus especially on VAC and HS impurities. We define $n$ $(0 \le n\le 1)$ as the fraction of the total impurity density $n_{\text{imp}}$ comprised of VAC in the case of vacancies, or HS in the case of adatoms, with $(1-n)$ the corresponding complementary fraction of REC or TOP impurities. The impurity mixture can be introduced into the transport formalism by writing the self energy as, e.g., $\Sigma(\omega^\pm) = n\Sigma_{H}(\omega^\pm) + (1-n)\Sigma_T(\omega^\pm)$, for a mixture of TOP and HS impurities. This is a good approximation in the dilute limit. We will study the four possible cases of single impurity species through the limit cases $n \rightarrow 0$ and $n \rightarrow 1$.

Calculating the coefficients $A^{\mu\mu}_{11}(\phi)$ and $A^{\mu\mu}_{22}(\phi)$, expanding the couplings about the symmetry points, and evaluating the momentum integrals, we obtain
\begin{widetext}
\begin{subequations}\label{eq:resistivity_final}
\begin{equation}
    \bar{\rho}_{\text{R-V}}(T,\mu) = \left(\frac{2S}{h n_{\text{imp}}}\frac{e^2}{h}\int_{-D}^{D}\ud \ee \left[ -\frac{\partial n_F(\ee,T,\mu)}{\partial \ee}\right]\frac{|\ee|}{\left[\frac{V^2}{2}  + \frac{V^2}{6t^2}\ee^2 \right](1-n)\rho_{dR}(\ee) + \frac{V^2}{t^2} \ee^2 n\rho_{dV}(\ee) }\right)^{-1},
\end{equation}
\begin{equation}
    \bar{\rho}_{\text{T-H}}(T,\mu) = \left(\frac{2S}{h n_{\text{imp}} }\frac{e^2}{h}\int_{-D}^{D}\ud \ee \left[ -\frac{\partial n_F(\ee,T,\mu)}{\partial \ee}\right]\frac{|\ee|}{\sqrt{\frac{V^2}{2} (1-n)\rho_{dT}(\ee)}\sqrt{\frac{V^2}{2} (1-n)\rho_{dT}(\ee) + 2( V/t )^2 \ee^2 n\rho_{dH}(\ee) }}\right)^{-1},
\end{equation}
\end{subequations}
\end{widetext}
where the Planck constant has been reintroduced. To work exclusively in the Dirac regime while also preserving the total number of states in the Brillouin zone, we have defined the Debye half--bandwidth\cite{peres_prb_2006} $D =  3^{-3/4}(8\pi)^{1/2}a^{-1}\hbar v_F \sim 10 \,\text{eV}$.  The problem is now reduced to calculating the single--impurity spectral densities $\rho_{dI}(\ee)$.

\section{Resistivity calculations}\label{sec:resistivity}
\subsection{The impurity spectral density}\label{sec:rhod}
\begin{figure}[t]
\begin{center}
\includegraphics[width=0.99\columnwidth]{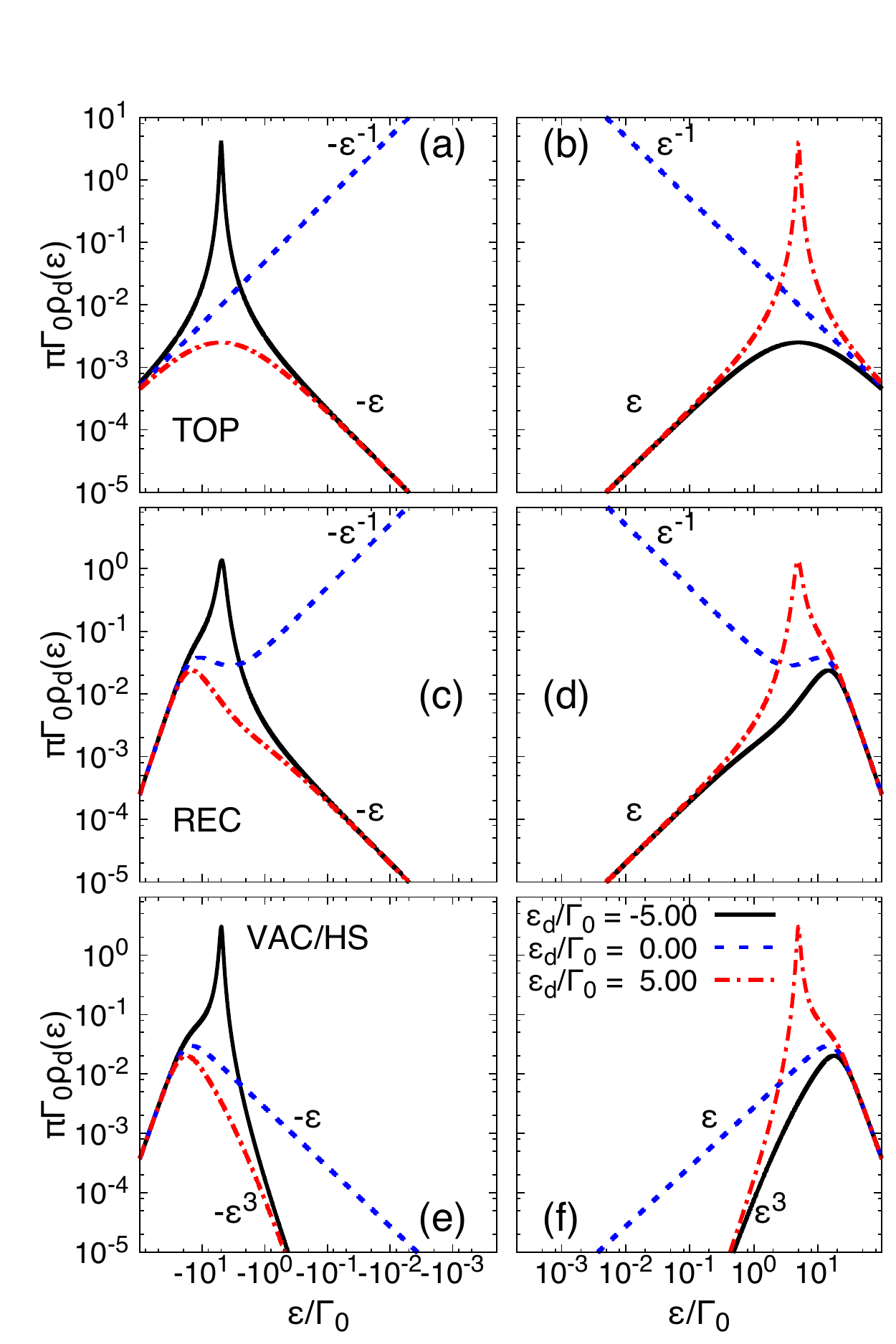}
\caption{(Color online) The spectral densities of TOP, REC, VAC and HS impurities for different local energies $\varepsilon_d$, using $\Gamma_0/D = 0.05$. The amplitude at $\varepsilon=0$ vanishes as a power law for the $C_{3v}$ impurities VAC and HS, in stark contrast to the non-$C_{3v}$ impurities REC and TOP, whose spectral densities are singular at zero energy.}
\label{fig:spectral_density}
\end{center}
\end{figure}

As stated in Eq.\ (\ref{eq:Himp}), we treat the impurities as single non--interacting orbitals of energy $\varepsilon_d$. The impurity spectral density (Fig.\ \ref{fig:spectral_density}) can be obtained by solving the equation of motion of the retarded Green's function:
\begin{equation}\label{eq:spectral_density}
    \rho_{dI}(\varepsilon) = -\frac{1}{\pi}\imag{\rgd{\varepsilon^+}} = \frac{1}{\pi}\frac{\Gamma_I(\varepsilon)}{[\varepsilon - \varepsilon_d - \Lambda_I(\varepsilon)]^2 + \Gamma_{I}^2(\varepsilon)}.
\end{equation}
The hybridization function for impurity type $I$ is defined as
\begin{equation}\label{eq:hyb}
    \Gamma_I(\ee) = \pi\sum_{\mu=1}^4\sum_{\vec{k}}|\Theta_I^\mu|^2\delta(\ee - \epsilon_{\mu}(k)),
\end{equation}
with 
$\epsilon_{1}(k) = \epsilon_{4}(k) = -\epsilon_{2}(k) = -\epsilon_{3}(k) = \hbar v_Fk$, 
and it can be shown that the level shifts $\Lambda_I(\varepsilon)$ vanish. We have
\begin{subequations}\label{eq:hyb_all}
\begin{equation}\label{eq:hyb_top}
    \Gamma_{T}(\ee) = \frac{2\pi V^2}{D^2}|\ee|\equiv \Gamma_0\left|\frac{\ee}{D}\right|,
\end{equation}
\begin{equation}\label{eq:hyb_rec}
    \Gamma_{R}(\ee) = \Gamma_0 \left[ \left|\frac{\ee}{D}\right| + 6\pi\sqrt{3}\left|\frac{\ee}{D}\right|^3 \right],
\end{equation}
\begin{equation}\label{eq:hyb_c3v}
    \Gamma_H(\ee) = \Gamma_V(\ee) =  4\pi\sqrt{3}\Gamma_0\left|\frac{\ee}{D}\right|^3.
\end{equation}
\end{subequations}
The dependence on the third power of the energy in the last expression led the authors of Ref.\ [\onlinecite{uchoa_prl_2011}] to predict super--ohmic transport through vacancies, hollow--site impurities and substitutional atoms. However, our analysis of Section \ref{sec:linear_response} and the results that we present in the following section demonstrate that this is not reflected in the transport properties of the graphene sample itself.

\subsection{Results and discussion}\label{sec:resultsdiscussion}
\begin{figure}[t]
\begin{center}
\includegraphics[width=0.99\columnwidth]{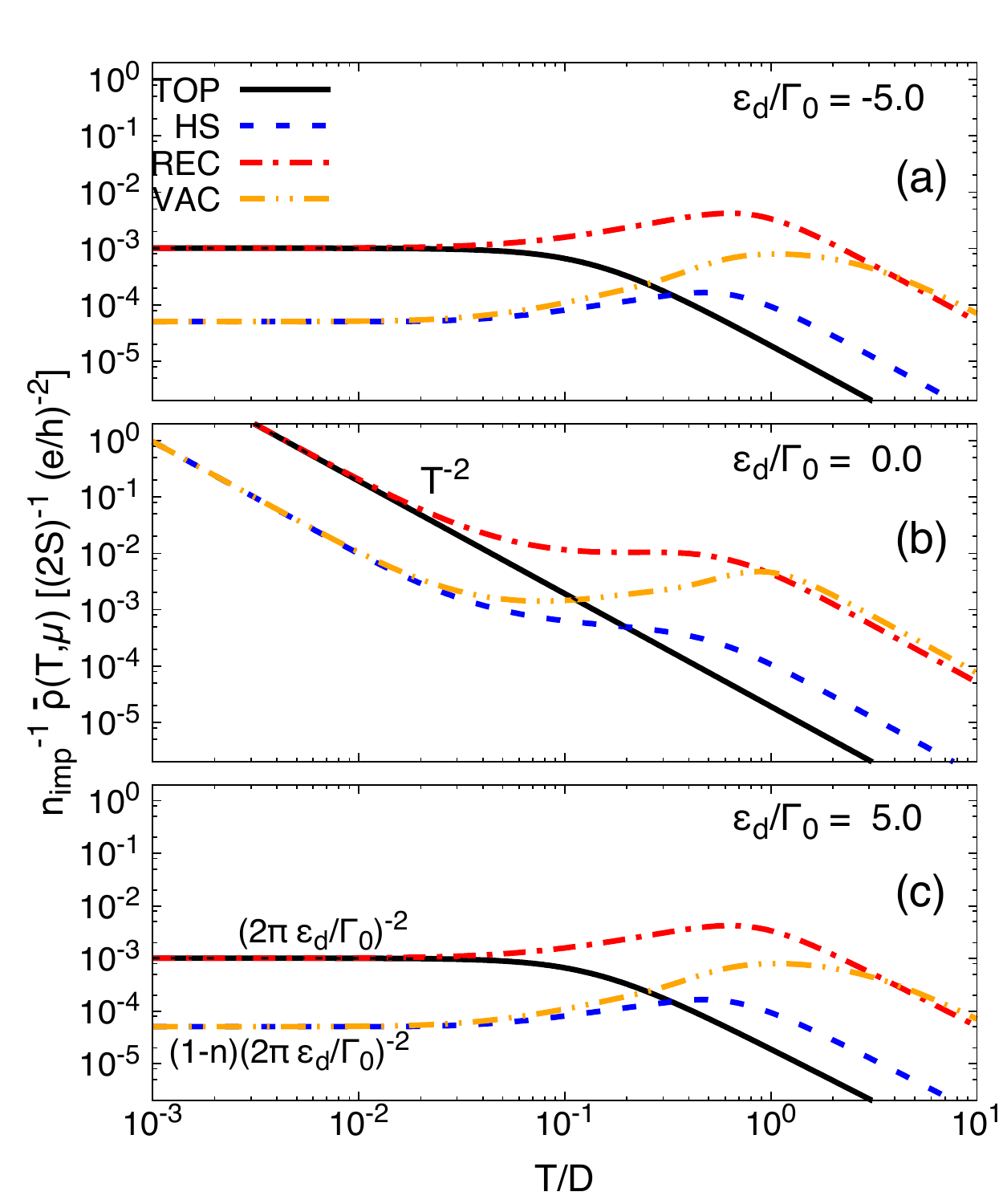}
\caption{(Color online) Resistivity as a function of temperature for TOP and REC ($n = 0$), and VAC and HS impurities ($n = 0.95$). Results are shown in charge neutrality ($\mu = 0$) for impurity local energies (a) $\varepsilon_d = -5\Gamma_0$, (b) $\varepsilon_d = 0$, and (c) $\varepsilon_d = 5\Gamma_0$, with $\Gamma_0 = 0.05\,D \sim 0.5 \mathrm{eV}$.}
\label{fig:res_vs_temp}
\end{center}
\end{figure}
The resistivity of a graphene sample in charge neutrality as a function of temperature is shown in Fig.\ \ref{fig:res_vs_temp} for all impurity types. The curves for VAC and HS impurities were obtained using $n=0.95$ in Eqs.\ (\ref{eq:resistivity_final}a) and (\ref{eq:resistivity_final}b), respectively. The corresponding value used for TOP and REC was $n=0$.

In general, the low--temperature resistivity contributions of TOP and REC impurities can be understood in terms of impurity scattering for all values of $\ee_d$. In all cases shown in Fig.\ \ref{fig:res_vs_temp} the curves for TOP and REC merge at temperatures below $|\ee_d|$, as one would anticipate given that the spectral densities of both impurity types are identical at low energies [Figs.\ \ref{fig:spectral_density}(a)-(d)]. For impurities off resonance with the Dirac point ($\ee_d \ne 0$) the resistivity reaches a saturation value of $(2\pi\,\varepsilon_d / \Gamma_0)^{-2}$, producing a plateau at low temperatures [Figs.\ \ref{fig:res_vs_temp}(a) and (c)]. This is a clear signature of impurity scattering dominating the electronic transport.

A substantial difference can be seen for TOP and REC impurities in resonance with the Dirac point ($\ee_d = 0$). In this case, the impurity introduces a bound state at zero energy\cite{ferreira_prb_2011,asmar_prl_2014} that dictates the low--energy behavior of the system. For TOP impurities, the resistivity is given by
\begin{equation}\label{eq:tm2}
    \bar{\rho}_{T}(T,\ee_d=\mu=0) = \left[\frac{2Se^2}{n_{\text{imp}}h^2}\,\frac{4\pi^4}{3}\frac{D^2 + \Gamma_0^2}{\Gamma_0^2}  \right]^{-1}\frac{D^2}{T^2},
\end{equation}
throughout the full range of temperatures [Fig.\ \ref{fig:res_vs_temp}(b)]. This function diverges as $T^{-2}$ for $T \rightarrow 0$---an insulating behavior
reflecting the suppression of thermally--activated transport as the temperature is lowered. Although REC impurities display the same behavior for $T < |\ee_d|$, the cubic energy term in Eq.\ (\ref{eq:hyb_rec}) dominates at high temperatures, producing a departure from the $T^{-2}$ scaling.

The curves for VAC and HS in Fig.\ \ref{fig:res_vs_temp} confirm that these impurities do not contribute to the resistivity of a graphene sample in charge neutrality, as discussed in Section \ref{sec:linear_response}. For $\ee_d \ne 0$ the low--temperature resistivity is given by $(1-n)(2\pi\,\varepsilon_d / \Gamma_0)^{-2}$ [Figs.\ \ref{fig:res_vs_temp}(a) and (c)], indicating that the resistivity originates from low--energy scattering with the small fraction (0.05) of either REC or TOP impurities present in the sample. Similarly, for $\ee_d = 0$ both curves follow a $T^{-2}$ trend at low temperatures, described by Eq.\ (\ref{eq:tm2}) with a prefactor $(1-n)$.

\begin{figure*}[t]
\begin{center}
\includegraphics[width=1.99\columnwidth]{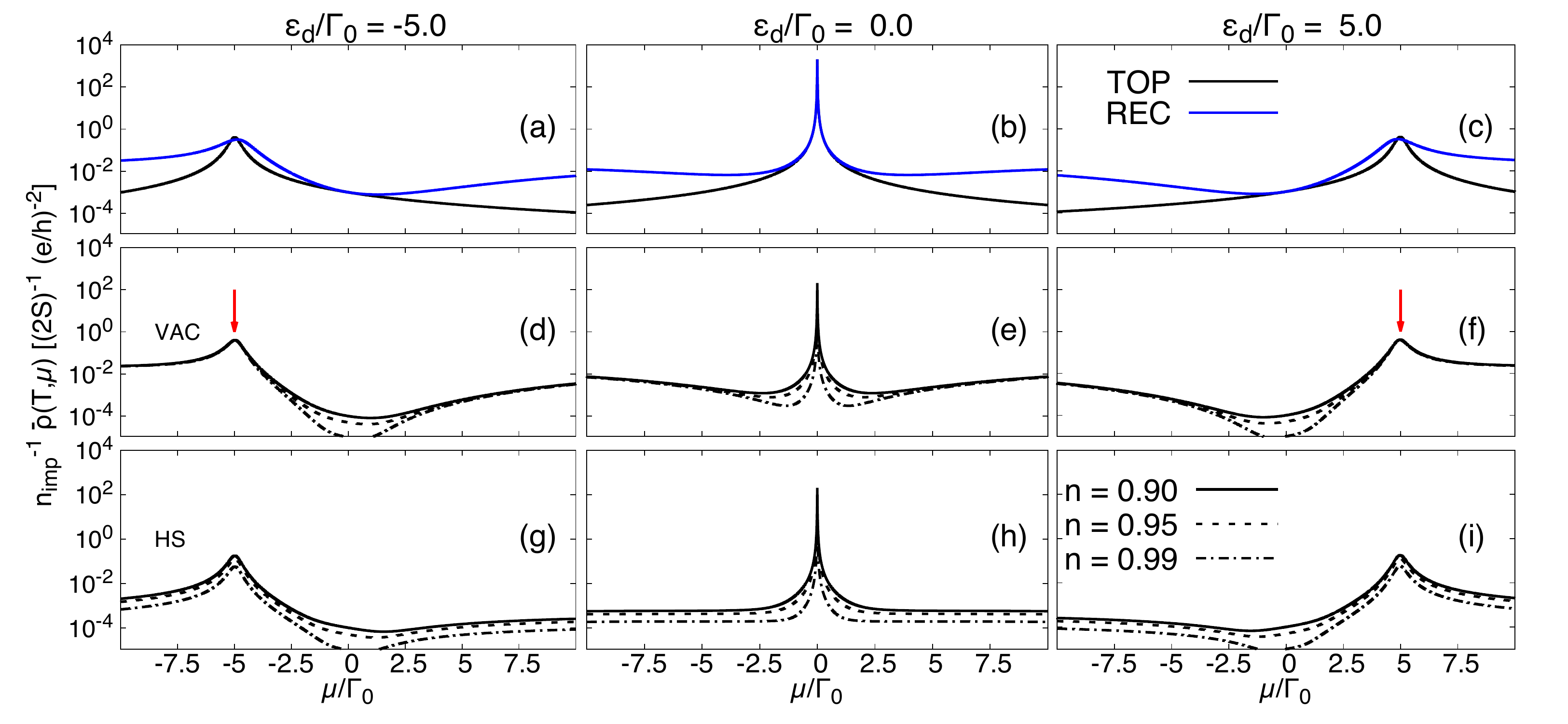}
\caption{(Color online) Low--temperature resistivity as a function of chemical potential. Panels (a), (b) and (c) correspond to a graphene sample with only 
(non--$C_{3v}$--symmetric) TOP and REC impurities. Panels (d) through (i) correspond to a mixture with a fraction $n$ of $C_{3v}$--symmetric and $1-n$ of non--$C_{3v}$--symmetric impurities. Results are shown for impurity local energies $\varepsilon_d = -5\Gamma_0,\,0,\,5\Gamma_0$ (left to right) at a temperature $T= 10^{-4}D \sim 10\,\mathrm{K}$. We have used $\Gamma_0 = 0.05\,D$, which gives an impurity-graphene coupling $V \approx 800\,\mathrm{meV}$.}
\label{fig:res_vs_mu}
\end{center}
\end{figure*}
Next, we examine the resistivity as a function of the chemical potential. Figure \ref{fig:res_vs_mu} shows isotherms at $T = 10^{-4}\,D \sim 10\,\mathrm{K}$ for all four impurity types. The figure shows curves for VAC and HS impurities using different values of $n$. In general terms, the resistivity of a sample with only TOP impurities [Figs.\ \ref{fig:res_vs_mu}(a)-(c)] has a maximum amplitude when $\mu = \varepsilon_d$ and impurity scattering is enhanced. For $\ee_d = 0$ the resistivity peak [Fig.\ \ref{fig:res_vs_mu}(b)] follows our previous discussion for charge neutrality, and increases as the temperature is lowered, following Eq.\ (\ref{eq:tm2}). The peak amplitude increase with decreasing temperature shown in Fig.\ \ref{fig:res_vs_mu}(b) is reminiscent of early resistivity measurements in graphene.\cite{novoselov_science_2004} For $\ee_d \ne 0$, on the other hand, the resistivity maximum has a low--temperature saturation value of $(2\pi\,\ee_d/D)^{-2}$ [Figs.\ \ref{fig:res_vs_mu}(a) and (c)]. The behavior is qualitatively the same for REC impurities, with a slightly shifted maximum and the appearance of a local minimum due to the cubic term in Eq.\ (\ref{eq:hyb_rec}), as can be seen in Figs.\ \ref{fig:res_vs_mu}(a) through \ref{fig:res_vs_mu}(c).

As before, the resistivity profiles for VAC and HS impurities are similar to those of REC and TOP impurities, respectively. Nonetheless, two important differences arise when varying the impurity fraction $n$: First, as the TOP or REC fraction goes to zero ($n \rightarrow 1$) the resistivity in charge neutrality vanishes for samples with only HS or VAC impurities, respectively. This behavior is independent of $\ee_d$ [Figs.\ \ref{fig:res_vs_mu}(d)-(i)]. Secondly, an important distinction appears between VAC and HS impurities away from charge neutrality for all values of $\ee_d$. While for hollow--site adatoms the resistivity goes to zero with ($1-n$) for all values of the chemical potential [Figs.\ \ref{fig:res_vs_mu}(g)-(i)], symmetric vacancies show a finite resistivity away from charge neutrality, even as $n$ goes to one. This is especially clear in the maxima indicated with arrows in Figs.\ \ref{fig:res_vs_mu}(d) and \ref{fig:res_vs_mu}(f), which have a fixed finite value independent of $n$.

\begin{figure}[h]
\begin{center}
\includegraphics[width=0.99\columnwidth]{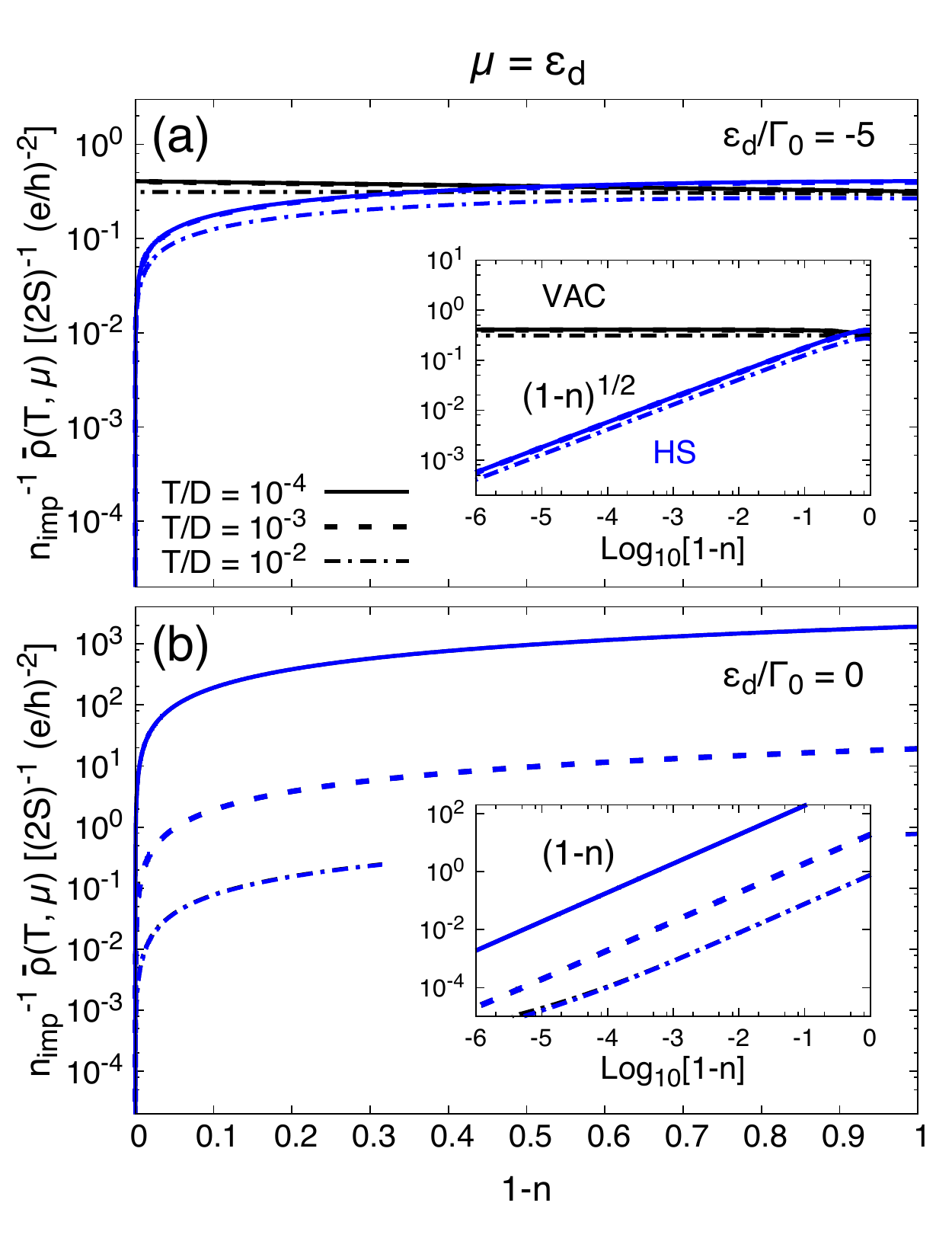}
\caption{(Color online) Resistivity of adatoms and vacancies as a function of the fraction $n$ of $C_{3v}$ impurities (HS and VAC) for fixed chemical potential $\mu = -5\,\Gamma_0$. (a) For vacancies, the case of $\mu = \ee_d = -5\,\Gamma_0$ corresponds to the resistivity maximum of Fig.\ \ref{fig:res_vs_mu}(d), which is independent of $n$. For adatoms this is also the maximum peak, but its amplitude goes to zero as $(1-n)^{1/2}$ as $n$ goes to one. (b) In the case of a resonant impurity ($\ee_d = 0$) the resistivity for adatoms also vanishes as $(1-n)^{1/2}$, whereas for vacancies it remains finite (away from charge neutrality) up to $n = 1$.}
\label{fig:res_vs_nv}
\end{center}
\end{figure}
Figure \ref{fig:res_vs_nv}(a) shows the behavior of the resistivity maxima of Figs.\ \ref{fig:res_vs_mu}(d) and \ref{fig:res_vs_mu}(g) as a function of the VAC and HS impurity fraction $n$. As described above, for symmetric vacancies the peak amplitude remains unchanged for all values of $n$, including the case of pure VAC impurities (inset). For adatoms, on the other hand, the corresponding amplitude goes to zero as $(1-n)^{1/2}$ for low HS fractions.  In the case of charge neutrality, however, the behaviors of VAC and HS impurities with $n$ are identical, as shown in Fig.\ \ref{fig:res_vs_nv}(b). Their corresponding resistivity curves overlap and go to zero as $(1-n)$. Notice that away from charge neutrality [Fig.\ \ref{fig:res_vs_nv}(a)] the resistivity for both VAC and HS impurities is quite independent of temperature, in agreement with experiments.\cite{novoselov_nature_2005,zhang_nature_2005}

The restoration of a finite resistivity contribution for VAC impurities is achieved by avoiding the graphene states at the Dirac points, which are impervious to the presence of symmetric vacancies. At a finite chemical potential, only states within an energy window $\sim T$ about the Fermi level will partake in transport processes. Therefore, coherent transport can be avoided for VAC impurities by shifting the chemical potential away from the charge neutrality point. However, this is not the case for HS impurities, which have momentum states available for coherent transport at all energies, and cannot be avoided by changing the carrier density. 

As discussed in Section \ref{sec:formalism}, this fundamental difference between the two highly--symmetric impurity types can be traced back to their distinct behaviors under inversion. While VAC impurities break it locally, HS impurities fully preserve the graphene inversion symmetry. Thus, while the $C_{3v}$ symmetry protects the vanishing of the coupling only at the symmetry points, inversion symmetry guarantees the appearance of nodes in the coupling function $\Theta_H(\vec{k})$ throughout the Brillouin zone.

The appearance of these nodes due to inversion symmetry can be seen as follows: A generic inversion--invariant hollow--site impurity couples to the graphene states as (Appendix \ref{app:inversion})
\begin{equation}
    |\Theta_H^\pm(\vec{k})|^2 = |V_{\vec{k}}|^2\left(1 \pm \real{\left[\exp{i\{\arg{\Phi(\vec{k})} + 2\arg{V_{\vec{k}}}\}} \right]}\right),
\end{equation}
where $V_{\vec{k}}=\sum_{j=1}^3V_j^*\exp{ia\vec{k}\cdot\hat{\vec{u}}_j}$, and $V_j$ is the real--space coupling to the sublattice--$B$ site at $a\hat{\vec{u}}_j$. Both terms between parentheses are of norm unity, such that the coupling will vanish for all  momenta $\vec{k}$ in subband $\alpha=\pm$ that fulfill,
\begin{equation}\label{eq:ccond_nodes}
    \arg{\Phi(\vec{k})} +2 \arg{V_{\vec{k}}} = [2n+(1\pm1)/2]\pi,
\end{equation}
with $n$ some integer. For simplicity, let us assume that $V_j$ are real, as in the particular case of Eq.\ (\ref{eq:vHS}). Then, after a few manipulations, Eq.\ (\ref{eq:ccond_nodes}) becomes
\begin{equation}
\begin{split}
    \sum_{j=1}^3\sum_{l=1}^3&V_jV_l\Big\{\sin(a\vec{k}\cdot[\hat{\vec{u}}_j + \hat{\vec{u}}_l])\real{\Phi(\vec{k})}\\ &+ \cos(a\vec{k}\cdot[\hat{\vec{u}}_j + \hat{\vec{u}}_l])\imag{\Phi(\vec{k})} \Big\} = 0.
\end{split}
\end{equation}
The appearance of the real and imaginary parts of $\Phi(\vec{k})$ guarantees that the symmetry points, where $\Phi(\vec{k})$ vanishes, are always solutions to this equation.

In fact, a numerical study of $\Theta_H(\vec{k})$ for different $V_j$ demonstrates that each symmetry point will be the termination of at least one node, and that for most cases of interest this node will be a straight line, as in the cases of Figs.\ \ref{fig:thetasq}(c) and \ref{fig:thetasq}(d).
\section{Conclusions}\label{sec:conclusions}
In this article we have studied the linear electronic transport properties of mesoscopic graphene with a low concentration of adatoms or vacancies. Our results for different impurities demonstrate distinct transport behaviors that can be traced to the point symmetry of their couplings to the graphene sublattices, and to whether they locally break or preserve the inversion symmetry of the honeycomb lattice.

Top adatoms and reconstructed vacancies, which break inversion symmetry locally, and whose symmetries differ from the sublattice $C_{3v}$ point--group, show similar behaviors. For a charge--neutral graphene sample, the usual impurity scattering contribution to the resistivity is present when the impurity is off resonance with the Dirac point. For a resonant (zero--energy) impurity, transport is strongly suppressed at low temperatures. In this case, the presence of a bound state at zero energy leads to a power--law divergence of the impurity resistivity in the zero--temperature limit. Away from charge neutrality, transport is again dominated by impurity scattering, which is maximized when the chemical potential is tuned to match the impurity level energy.

More interesting is the behavior of hollow--site ($s$-level) adatoms and $C_{3v}$--symmetric vacancies. Their contributions to the resistivity of a charge--neutral graphene sample \textit{vanish} due to the presence of electronic states that are fully decoupled from the impurities, and thus impervious to their presence. 

In the case of symmetric vacancies, which locally break inversion symmetry, the decoupled states correspond exactly to those located at the $K$ and $K'$ symmetry points. The contribution by these states to the transport can be prevented by changing the chemical potential through, e.g., the application of a gate voltage, and thus a finite resistivity is recovered. 

This is not the case for hollow--site adatoms, which decouple from full line nodes of momentum states all throughout the Brillouin zone. The specific momenta forming these nodes depend on the particular rotational symmetry of the impurity, but their existence is protected by the local conservation of inversion symmetry.

We believe it should be possible to verify our predictions through standard transport measurements on gated graphene samples. For an estimated nearest--neighbor hopping $t = 2.7 \, \mathrm{eV}$, the results presented in Figs.\ \ref{fig:res_vs_mu} and \ref{fig:res_vs_nv} correspond to temperatures of order $10\,\mathrm{K}$ or higher, and a realistic\cite{wehling_prl_2010,yuan_prb_2010} impurity coupling $V \approx 800\, \mathrm{meV}$.

Our results can be readily generalized for interacting impurities and vacancies in graphene. The formation of local magnetic moments may be introduced by considering a local Coulomb interaction term in Eq.\ (\ref{eq:Himp}). Such magnetic impurities---predicted for transition--metal\cite{eelbo_prl_2013} and hydrogen\cite{mccreary_prl_2012} adatoms, and both symmetric and reconstructed vacancies\cite{yazyev_prb_2007,Cazalilla12,Miranda2016,rodrigo_carbon_2016}---would introduce strong correlations that can be handled by the numerical renormalization group (NRG). This non--perturbative method can correctly evaluate the interacting impurity spectral density entering Eqs.\ (\ref{eq:resistivity_final}), and thus the resulting resistivity contribution.

For charge--neutral graphene, in particular, the interacting impurity problem can be described by the so-called pseudogap Anderson model. \cite{VojtaEPL:27006:2010,fritz_vojta_kondo_in_graphene} In this case, the interacting--impurity spectral density vanishes at the Fermi level with the same a power law as the density of states, \cite{Dias:153304:2008} similarly to the results presented in Fig.\ \ref{fig:spectral_density}. To put it differently, the local density of states for interacting impurities in graphene will vanish at the Fermi energy as $\rho(\omega) \sim |\omega|$ for symmetry--breaking impurities, and as $\rho(\omega) \sim |\omega|^3$ for symmetry--preserving ones, just like in the non--interacting case. 
However, in the presence of long--range disorder\cite{Miranda14} or a finite chemical potential, we expect the low--temperature behavior of the system to be dominated by Kondo correlations. Although the latter case has been discussed by some authors,\cite{cornaglia_prl_2009,kanao_jpsj_2012,po-wei-lo_prb_2014} the momentum dependence of the impurity--graphene coupling due to symmetry has yet to be addressed.

\textit{Note added:} After the completion of this work, we became aware of a manuscript \cite{ferreira_prb_2016} which has studied the impurity scattering properties for the cases of top-site and hollow-site adatoms. Their conclusions are in accordance to our results, namely that impurity scattering is strongly suppressed in the case of hollow--site impurities.

\begin{acknowledgments}
The authors thank Caio Lewenkopf and Tatiana Rappoport for enlightening discussions and suggestions. D.A.R.T.\  thanks Mahmoud Asmar for many fruitful discussions during the preparation of this article, and Aires Ferreira for useful comments on the transport formalism. D.A.R.T.\  acknowledges financial support by the Brazilian agency CAPES. L.G.G.V.D.S. acknowledges financial support by CNPq (grants No. 307107/2013-2 and 449148/2014-9), PRP-USP NAP-QNano and FAPESP.
\end{acknowledgments}

\appendix
\section{Real--space impurity--graphene couplings }\label{app:impurities}
Here we present the expressions for the impurity-graphene couplings $\Hig$ in real space for the different impurity types of impurities ($I=T,\,H\,,V,\,R$). Without loss of generality we set the origin of our coordinate system at the impurity site. When the impurity sits at or on top of a lattice site, as is the case for TOP, VAC and REC, we call the corresponding sublattice $A$.

TOP impurities couple to a single site as
\begin{equation}\label{eq:topA}
    \HtopA = V\sum_{s}\left\{d_s^\dagger a_s(0) + a_s^\dagger(0) d_s \right\}.
\end{equation}
A VAC impurity will couple identically to all three surrounding sublattice $B$ sites located at $a\uu_j$ as
\begin{equation}\label{eq:vacA}
    \HvacA = V\sum_{s}\sum_{j=1}^3\left\{ d_s^\dagger b_s(a\uu_j) + \text{H.\ c.}  \right\}
\end{equation}
For the case of an asymmetric REC impurity, we consider that the $sp^2$ orbital of the $B$--sublattice carbon atom at $a\hat{\vec{u}}_l$ will couple to the $\pi$ orbitals of the two $B$--sublattice carbons at $a\hat{\vec{u}}_j$ ($\l \ne j$) as
\begin{equation}\label{eq:recA}
    \HrecA(l) = V\sum_{s}\sum_{j=1}^3(1-\delta_{j,l})\left\{ d_s^\dagger b_s(a\uu_j - a\uu_l) + \text{H.\ c.}  \right\}
\end{equation}
Finally, HS impurities couple identically to both sublattices:
\begin{equation}\label{eq:hs}
    \Hhs =V\sum_{s}\sum_{j=1}^3\big\{ d_s^\dagger \big[a_s(a\uu_j) + b_s(-a\uu_j)\big] + \text{H.\ c.} \big\}
\end{equation}
In Fourier space we have
\begin{subequations}
\begin{equation}
    \HtopA = V\sum_{\vec{k},s}\left\{d_s^\dagger a_{\vec{k}s} + \text{H.\ c.}  \right\},
\end{equation}
\begin{equation}
    \HvacA = V\sum_{\vec{k},s}\left\{\Phi(\vec{k})d_s^\dagger b_{\vec{k}s} + \text{H.\ c.}  \right\},
\end{equation}
\begin{equation}
    \HrecA(l) = V\sum_{\vec{k},s}\left\{\exp{-i\vec{k}\cdot\uu_l}\left[\sum_{j\ne l}\exp{ia\vec{k}\cdot\uu_j}\right] d_s^\dagger b_{\vec{k}s} + \text{H.\ c.}  \right\},
\end{equation}
\begin{equation}
\begin{split}
    \Hhs =&V\sum_{s}\big\{ d_s^\dagger \big[\Phi(\vec{k})a_{\vec{k}s} + \Phi^*(\vec{k})b_{\vec{k}s}\big] + \text{H.\ c.} \big\}.
\end{split}
\end{equation}
\end{subequations}
Applying the transformation (\ref{eq:unitary_2band}) we obtain Eqs.\ (\ref{eq:vVAC}), (\ref{eq:vREC}), (\ref{eq:vTOP}),  and (\ref{eq:vHS}).

\section{Kubo formula for the zero--bias conductivity}\label{app:kubo}
\begin{figure}[h]
\begin{center}
\includegraphics[width=0.85\columnwidth]{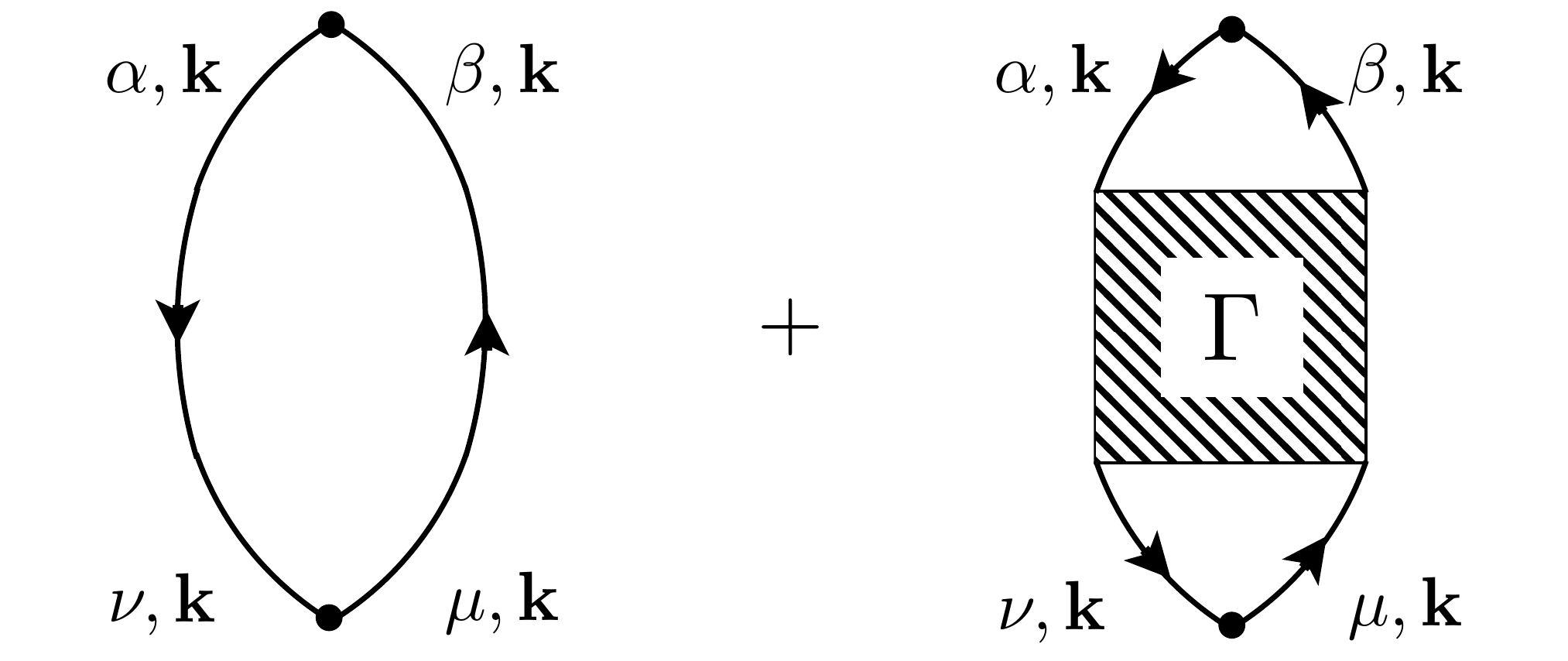}
\caption{The current-current correlation function is represented by a polarization diagram and an interaction vertex $\Gamma$. Lines represent full graphene Green's functions in the presence of the impurities, and point vertices at the top and bottom represent the current operators.}
\label{fig:diagrams}
\end{center}
\end{figure}
In the linear response regime, the electric conductivity tensor is obtained  through the Kubo formula
\begin{equation}\label{eq:cond_response}
    \sigma^{ij} = \lim_{\omega \rightarrow 0}\frac{1}{i\omega}\imag{\Pi^{ij}(\omega^+)},
\end{equation}
where $\Pi^{ij}(\omega^+)$ is the retarded response function, and $\omega$ the (angular) frequency of the driving electric field. Following common practice, we work in the imaginary time and frequency domain to simplify the calculations. The Matsubara response function is given by
\begin{equation}\label{eq:response_matsubara}
    \Pi^{ij}(i\omega_n) = \int_0^{1/T} \ud \tau\,\exp{i\omega_n\tau}\xpect{j^i(\tau)j^j(0)},
\end{equation}
where $\omega_n = 2 n \pi T$ is a bosonic Matsubara frequency, $j^i(\tau)$ is the $i$th component of the imaginary--time Heisenberg representation of the zero--momentum current operator
\begin{equation}
    \vec{j}(\tau) \equiv \exp{H\tau}\vec{j}(\vec{q}=0)\exp{-H\tau},
\end{equation}
and $T$ is the temperature (in units of energy). We can relate this quantity to the retarded response function through the formula
\begin{equation}
    \Pi^{ij}(\omega^+) = -T^{-1}\, \Pi^{ij}(i\omega_n \rightarrow \omega + i0^+).
\end{equation}

The two--point current correlation function can be evaluated in the graphene basis using Eq.\ (\ref{eq:current_final}):
\begin{equation}
\begin{split}
    \Pi^{ij}(i\omega_n) = (ev_F)^2\sum_{\vec{k},\vec{k}'}&\int_0^{1/T} \ud \tau\,\exp{i\omega_n\tau}\gamma_{\alpha\beta}^i(\vec{k})\gamma_{\mu\nu}^j(\vec{k}')\\ &\times \xpect{c_{\vec{k}}^{\alpha}{}^\dagger(\tau) c_{\vec{k}}^\beta(\tau) c_{\vec{k}'}^\mu{}^\dagger c_{\vec{k}'}^\nu},
\end{split}
\end{equation}
where a sum is implied over all repeated Greek indices. Using Wick's theorem\cite{wick_1950} to expand the correlation function, and performing a partial summation of the resulting diagrams, the response function is represented by the two diagrams of Fig.\ \ref{fig:diagrams}. In the ``ladder'' approximation,\cite{bickers_1987} the vertex $\Gamma$ contributes an overall factor of order one to the full propagator result. Since no qualitative changes are introduced, we retain only the ``bubble'' diagram representing the expression
\begin{equation}
\begin{split}
    \Pi^{ij}(i\omega_n) = -(ev_F)^2\sum_{\vec{k}}\int_0^{1/T} \ud \tau\,&\exp{i\omega_n\tau}\gamma^i_{\alpha\beta}(\vec{k}) \gamma^j_{\mu\nu}(\vec{k})\\ &\times\gs{\vec{k}\vec{k}}{\nu\alpha}{-\tau}\gs{\vec{k}\vec{k}}{\beta\mu}{\tau}.
\end{split}
\end{equation}
At this point we introduce the Fourier representation of the imaginary--time Green's functions
\begin{equation}
    \gs{\vec{k}\vec{k}'}{\beta\mu}{\tau} = T\sum_{ik_m}\exp{-i\,k_m\tau}\gs{\vec{k}\vec{k}'}{\beta\mu}{ik_m},
\end{equation}
with $k_m=(2m + 1)\pi T$ a fermionic Matsubara frequency. The imaginary--time integral is now easily evaluated to obtain
\begin{equation}
\begin{split}
    \Pi^{ij}(i\omega_n) = -(ev_F)^2T\sum_{\vec{k},\,ik_m}&\gamma^i_{\alpha\beta}(\vec{k}) \gamma^j_{\mu\nu}(\vec{k})\\ \times &\gs{\vec{k}\vec{k}}{\nu\alpha}{i[k_m+\omega_n]}\gs{\vec{k}\vec{k}}{\beta\mu}{ik_m}.
\end{split}
\end{equation}
The Matsubara sum can be evaluated by standard techniques, and retaining only the dominant term\cite{mahan} we obtain
\begin{equation}\label{eq:response_final}
\begin{split}
    \imag{\Pi^{ij}(\omega^+)} =&-(ev_F)^2\sum_{\vec{k}}\int_{-\infty}^{\infty}\frac{\ud \omega'}{2\pi}\gamma_{\alpha\beta}^i(\vec{k})\gamma_{\mu\nu}^j(\vec{k}) \\ &\times\rgs{\vec{k}\vec{k}}{\nu\alpha}{\omega'{}^{+} + \omega}\rgs{\vec{k}\vec{k}}{\beta\mu}{\omega'{}^{-}}\\
    &\times\left[n_F(\omega'+\omega,T) - n_F(\omega',T) \right],
\end{split}
\end{equation}
where $n_F(\omega,T)$ is the Fermi-Dirac distribution and $\rgs{\vec{k}\vec{k}}{}{\omega^\pm}$ is the retarded (advanced) Green's function. Finally, the conductivity is obtained by substituting (\ref{eq:response_final}) into (\ref{eq:cond_response}):\cite{ryu_prb_2007}
\begin{widetext}
\begin{equation}\label{eq:conductivity_app}
\begin{split}
    \sigma^{ij}(T)=&\frac{(ev_F)^2}{2\pi}\sum_{\vec{k}}\int_{-\infty}^{\infty}\ud\omega \left[-\frac{\partial n_F(\omega,T)}{\partial \omega} \right]\gamma_{\alpha\beta}^i(\vec{k})\gamma_{\mu\nu}^j(\vec{k})\,\rgs{\vec{k}\vec{k}}{\nu\alpha}{\omega^+}\rgs{\vec{k}\vec{k}}{\beta\mu}{\omega^-}\\
    =&\frac{(ev_F)^2}{2\pi}\sum_{\vec{k}}\int_{-\infty}^{\infty}\ud\omega\left[-\frac{\partial n_F(\omega,T)}{\partial \omega} \right]\trace{\gamma^i(\vec{k})\RGs{\vec{k}\vec{k}}{\omega^-}\gamma^j(\vec{k})\RGs{\vec{k}\vec{k}}{\omega^+}}.
\end{split}
\end{equation}
\end{widetext}
This formula is used in Section \ref{sec:resultsdiscussion} to evaluate the resistivity tensor as $[\rho^{-1}(T)]^{ij} = \sigma^{ij}(T)$.

\section{The full Green's function and the effective self energies}\label{app:Seff}
In this appendix, we present the expressions for the Green's functions entering Eq.\ (\ref{eq:trace_general}) in the dilute impurity limit. The self energy is defined by the identity
\begin{equation}\label{eq:def_self_energy}
    [\RGs{\vec{k}\vec{k}}{\omega^\pm}]^{-1} = [\Gz{\vec{k}}{\omega^\pm}]^{-1} - \SE{\vec{k},\,\omega^\pm},
\end{equation}
with $\Gz{\vec{k}}{\omega^\pm}$ the bare graphene Green's function, given by
\begin{equation}
    \Gz{\vec{k}}{\omega^\pm} = \begin{pmatrix}
    \frac{1}{\omega^\pm - v_Fk} & 0 & 0 & 0\\
    0 & \frac{1}{\omega^\pm + v_Fk} & 0 & 0\\
    0 & 0 & \frac{1}{\omega^\pm + v_Fk} & 0\\
    0 & 0 & 0 & \frac{1}{\omega^\pm - v_Fk}
    \end{pmatrix}.
\end{equation}
Substituting Eqs.\ (\ref{eq:self_energy}) and (\ref{eq:t_matrix}) into (\ref{eq:def_self_energy}) and inverting the resulting matrix, we obtain an analytic expression for the full graphene Green's function in terms of the impurity local Green's function.

The $11$ and $22$ components are given to first order in $n_{\text{imp}}$ by
\begin{subequations}
\begin{equation}
    \rgs{\vec{k}\vec{k}}{11}{\omega^\pm} = \frac{\omega + v_F k - |\Theta^1(\vec{k})|^2n_{\text{imp}}\rgd{\omega^\pm}}{\omega^2-(v_F k)^2 - 2\omega|\Theta^1(\vec{k})|^2n_{\text{imp}}\rgd{\omega^\pm}},
\end{equation}
\begin{equation}
    \rgs{\vec{k}\vec{k}}{22}{\omega^\pm} = \frac{\omega - v_F k - |\Theta^2(\vec{k})|^2n_{\text{imp}}\rgd{\omega^\pm}}{\omega^2-(v_F k)^2 - 2\omega|\Theta^2(\vec{k})|^2n_{\text{imp}}\rgd{\omega^\pm}}.
\end{equation}
\end{subequations}
For $\omega^2 \ne (v_Fk)^2$ and $n_{\text{imp}}\ll 1$ the denominator can be expanded in a geometric series, and each of the above Green's functions can be written (no sum over $\mu$ implied)
\begin{equation}
    \rgs{\vec{k}\vec{k}}{\mu\mu}{\omega^\pm}=\gz{\vec{k}}{\mu}{\omega^\pm} + \gz{\vec{k}}{\mu}{\omega^\pm} \Teff{\mu\mu}{\omega^\pm} \gz{\vec{k}}{\mu}{\omega^\pm},
\end{equation}
where the effective $T$ matrices are given by $\Teff{11}{\omega^\pm} = n_{\text{imp}}|\Theta_I^1(\vec{k})|^2\rgd{\omega^\pm}$ and $\Teff{22}{\omega^\pm} = n_{\text{imp}}|\Theta^2_I(\vec{k})|^2\rgd{\omega^\pm}$. Using Eq.\ (\ref{eq:self_energy}) we can define the effective self energies
\begin{equation}
    \Seff{\mu\mu}{\omega^\pm} = n_{\text{imp}}|\Theta_I^\mu(\vec{k})|^2\rgd{\omega^\pm}.
\end{equation}

The resistivity can be evaluated in terms of $\Seff{\mu\mu}{\omega^\pm}$. At first glance, Eq.\ (\ref{eq:trace_general}) requires all terms $\rgs{\vec{k}\vec{k}}{\mu\mu}{\omega^-}\rgs{\vec{k}\vec{k}}{\nu\nu}{\omega^-}$ and $\rgs{\vec{k}\vec{k}}{\mu\nu}{\omega^-}\rgs{\vec{k}\vec{k}}{\nu\mu}{\omega^-}$. Upon further inspection, however, the products $\rgs{\vec{k}\vec{k}}{\mu\mu}{\omega^-}\rgs{\vec{k}\vec{k}}{\nu\nu}{\omega^+}$ for $\mu \ne \nu$ vanish identically. To show this, let us compute
\begin{widetext}
\begin{equation}
\begin{split}
    \rgs{\vec{k}\vec{k}}{11}{\omega^-}\rgs{\vec{k}\vec{k}}{22}{\omega^+}=& \frac{1}{\omega - v_Fk - \real{\Seff{11}{\omega^+}}+i\imag{\Seff{11}{\omega^+}}}\frac{1}{\omega + v_Fk - \real{\Seff{22}{\omega^+}}-i\imag{\Seff{22}{\omega^+}}}\\
    =& \frac{\omega - v_Fk - \real{\Seff{11}{\omega^+}}-i\imag{\Seff{11}{\omega^+}}}{\left[\omega - v_Fk - \real{\Seff{11}{\omega^+}}\right]^2+\left[\imag{\Seff{11}{\omega^+}}\right]^2}\frac{\omega + v_Fk - \real{\Seff{22}{\omega^+}}+i\imag{\Seff{22}{\omega^+}}}{\left[\omega + v_Fk - \real{\Seff{22}{\omega^+}}\right]^2+\left[\imag{\Seff{22}{\omega^+}}\right]^2}\\
    =&\frac{\pi \rho^{11}_{\vec{k}\vec{k}}(\omega)}{n_{\text{imp}}|\Theta_I^1(\vec{k})|^2\imag{\Seff{11}{\omega^-}}}\frac{\pi \rho^{22}_{\vec{k}\vec{k}}(\omega)}{n_{\text{imp}}|\Theta_I^2(\vec{k})|^2\imag{\Seff{22}{\omega^-}}}\\
    &\times \left[\omega - v_Fk - \real{\Seff{11}{\omega^+}}-i\imag{\Seff{11}{\omega^+}} \right]\left[ \omega + v_Fk - \real{\Seff{22}{\omega^+}}+i\imag{\Seff{22}{\omega^+}}\right],
\end{split}
\end{equation}
\end{widetext}
where $\rho_{\vec{k}\vec{k}}^{\mu\mu}(\omega) = \delta(\omega - \ee^\mu(k)) + \mathcal{O}(n_{\text{imp}})$ are graphene spectral densities. In the limit $n_{\text{imp}}\rightarrow 0$, the above expression is proportional to $ \delta(\omega - v_Fk) \delta(\omega + v_Fk)$ and vanishes identically.

Next, notice that when $n_{\text{imp}}\rightarrow 0$ the Green's functions $\rgs{\vec{k}\vec{k}}{\mu\nu}{\omega^\pm}$ with $\mu \ne \nu$, containing the inter--band and inter--valley processes, vanish. In other words, these Green's functions must be at least of order $n_{\text{imp}}$, and the product of the advanced and retarded functions must be at least $\mathcal{O}(n_{\text{imp}}^2)$. The terms $\rgs{\vec{k}\vec{k}}{\mu\mu}{\omega^-}\rgs{\vec{k}\vec{k}}{\mu\mu}{\omega^+}$, on the other hand, are given by
\begin{subequations}
\begin{equation}\label{eq:adv_ret_1}
    \rgs{\vec{k}\vec{k}}{11}{\omega^-}\rgs{\vec{k}\vec{k}}{11}{\omega^+} = \frac{\delta(\omega - v_Fk)}{n_{\text{imp}}|\Theta_I^1(\vec{k})|\rho_d(\omega)},
\end{equation}
\begin{equation}\label{eq:adv_ret_2}
    \rgs{\vec{k}\vec{k}}{22}{\omega^-}\rgs{\vec{k}\vec{k}}{22}{\omega^+} = \frac{\delta(\omega + v_Fk)}{n_{\text{imp}}|\Theta_I^2(\vec{k})|\rho_d(\omega)},
\end{equation}
\end{subequations}
where $\rho_d(\omega)= -\pi^{-1}\imag{\rgd{\omega^+}}$ is the impurity spectral density. These expressions, representing intra--band and intra--valley processes, are $\mathcal{O}\{ n_{\text{imp}}^{-1} \}$, and dominate in the dilute limit.

\section{An effective Hamiltonian for graphene with a low impurity density}\label{app:inversion}

Most discussions in the literature about impurities in graphene consider the Dirac approximation, where the graphene Hamiltonian is $H_{D} = \hbar v_F\tau^3\vec{\sigma}\cdot\vec{k}$. The impurities are then introduced through terms of the form $H_{\text{imp}}=\sum_{i,j}A_{ij}\tau^i\sigma^j$. This picture is particularly useful for discussing whether the graphene symmetries are preserved or broken by the impurities. For example, inversion symmetry is broken within a valley by terms proportional to $\sigma^3$.

To make a connection with this picture, we define an effective Hamiltonian $H_{\text{eff}}$ for the graphene sample with a dilute impurity distribution through
\begin{equation}\label{eq:gf_ab}
    \Delta_{\vec{k}\vec{k}}^{-1}(\omega^+) = \omega^+ - H_{\text{eff}}(\vec{k},\omega^+),
\end{equation}
where $\Delta_{\vec{k}\vec{k}}(\omega^+) = U_{\vec{k}}\RGs{\vec{k}\vec{k}}{\omega^+}U_{\vec{k}}^{-1}$ is the propagator in the ($2\times 2$) $\psi$-basis. The resulting model has the form
\begin{equation}\label{eq:heff}
    H_{\text{eff}} = \lambda\sigma^0 + \chi\sigma^3 -\tilde{t}\left[\real{\Phi(\vec{k})}\sigma^1 - \imag{\Phi(\vec{k})}\sigma^2 \right],
\end{equation}
with
\begin{widetext}
\begin{subequations}
\begin{equation}
    \tilde{t}(\vec{k},\omega^+) = -\frac{3atk + n_{\text{imp}}\RGs{d}{\omega^+}\left\{g_{\vec{k}}^1(\omega^+)[g_{\vec{k}}^2(\omega^+)]^{-1}|\Theta_I^{+}(\vec{k})|^2 + g_{\vec{k}}^2(\omega^+)[g_{\vec{k}}^1(\omega^+)]^{-1}|\Theta_I^{-}(\vec{k})|^2 +2i\imag{\Theta_I^{+}(\vec{k})\Theta_I^{-}{}^*(\vec{k})} \right\}}{\frac{3ak}{2} + 3ak\,n_{\text{imp}}\RGs{d}{\omega^+}\left[g_{\vec{k}}^1(\omega^+)|\Theta_I^{+}(\vec{k})|^2 + g_{\vec{k}}^2(\omega^+)|\Theta_I^{-}(\vec{k})|^2 \right]},
\end{equation}
\begin{equation}
    \lambda(\vec{k},\omega^+) = \omega^+ - \frac{[g_{\vec{k}}^1(\omega^+)]^{-1}+[g_{\vec{k}}^2(\omega^+)]^{-1}+n_{\text{imp}}\RGs{d}{\omega^+}\left\{g_{\vec{k}}^1(\omega^+)[g_{\vec{k}}^2(\omega^+)]^{-1}|\Theta_I^{+}(\vec{k})|^2 + g_{\vec{k}}^2(\omega^+)[g_{\vec{k}}^1(\omega^+)]^{-1}|\Theta_I^{-}(\vec{k})|^2 \right\}}{2 + 2n_{\text{imp}}\RGs{d}{\omega^+}\left[g_{\vec{k}}^1(\omega^+)|\Theta_I^{+}(\vec{k})|^2 + g_{\vec{k}}^2(\omega^+)|\Theta_I^{-}(\vec{k})|^2 \right]} ,
\end{equation}
\begin{equation}
    \chi(\vec{k},\omega^+) = -\frac{1}{2}\frac{\real{\Theta_I^{+}(\vec{k})\Theta_I^{-}{}^*(\vec{k})}n_{\text{imp}}\RGs{d}{\omega^+}}{1 + n_{\text{imp}}\RGs{d}{\omega^+}\left[g_{\vec{k}}^1(\omega^+)|\Theta_I^{+}(\vec{k})|^2 + g_{\vec{k}}^2(\omega^+)|\Theta_I^{-}(\vec{k})|^2 \right]}.
\end{equation}
\end{subequations}
\end{widetext}
The second term in Eq.\ (\ref{eq:heff}) breaks inversion symmetry proportionally to the impurity density, unless $\chi(\vec{k},\omega^+)$ vanishes for all momenta. In other words, the condition for the impurity to preserve inversion symmetry is
\begin{equation}\label{eq:cond_inv}
    \real{\left[\Theta_I^+(\vec{k})\Theta_I^-{}^*(\vec{k})\right]} = 0.
\end{equation}
Consider a generic HS impurity ($I = H$) that couples to the $A$ sublattice carbon atoms at $a\hat{\vec{u}}_j$ and to the $B$ sublattice atoms at $-a\hat{\vec{u}}_j$ about the impurity sites through complex couplings $V_j^A$ and $V_j^B$, respectively ($j=1,\,2,\,3$). For this case the condition (\ref{eq:cond_inv}) translates to
\begin{equation}\label{eq:ccond_inv}
    \real{\left\{\sum_{j=1}^3\sum_{l=1}^3(V_j^BV_l^{B*} - V_l^AV_j^{A*})\exp{ia\vec{k}\cdot(\hat{\vec{u}}_l - \hat{\vec{u}}_j)}\right\}} = 0.
\end{equation}
A sufficient condition for this is that $V_j^A = V_j^{B*}$, which is broader than the notion of inversion symmetry discussed throughout this article, where we have focused on the case of real $V_j^A$ and $V_j^B$.

Indeed, inversion symmetry can be preserved by more general impurity models. A simple example is the case of an in--plane $f$-level hollow--site impurity discussed in Ref.\ [\onlinecite{uchoa_prl_2011}], which couples to its surrounding carbon atoms with equal strength but alternating sign. This type of model corresponds to $V_j^A = V_j^{B*} =  i V$, resulting in a coupling function $\Theta_f^{\pm}(\vec{k})=iV[\Phi^*(\vec{k}) \mp \Phi^2(\vec{k})/|\Phi(\vec{k})|]/\sqrt{2}$. Comparing to Eq.\ (\ref{eq:vHS}) we can conclude that both impurity types display the same transport behavior.

For an impurity satisfying $V_j^A = V_j^{B*}$, the squared coupling strength $|\Theta_H^\pm(\vec{k})|^2$ is given by
\begin{equation}
    |\Theta_H^\pm(\vec{k})|^2 = |V_{\vec{k}}|^2 \pm \real{\left\{\frac{\Phi(\vec{k})}{|\Phi(\vec{k})|}V_{\vec{k}}^2\right\}},
\end{equation}
where $V_{\vec{k}}=\sum_{j=1}^3V_j^{B*}\exp{ia\vec{k}\cdot\hat{\vec{u}}_j}$. Given that $\Phi(\vec{k})/|\Phi(\vec{k})| = \exp{i\arg{\Phi(k)}}$ has modulus unity, the inversion symmetry condition makes it possible for $|\Theta_H^\pm(\vec{k})|^2$ to vanish. The functions
\begin{equation}
    |\Theta_H^\pm(\vec{k})|^2 = |V_{\vec{k}}|^2\left(1 \pm \real{\left[\exp{i\{\arg{\Phi(\vec{k})} + 2\arg{V_{\vec{k}}}\}} \right]}\right),
\end{equation}
have zeroes for momenta $\vec{k}$ such that
\begin{equation}\label{eq:arg}
    \arg{\Phi(\vec{k})} + 2\arg{V_{\vec{k}}} = [2n+(1\pm1)/2]\pi,
\end{equation}
with $n$ an integer. These momenta are determined by the spatial symmetry of the coupling, but notice that no particular symmetry requirements are placed on $V_{\vec{k}}$ for (\ref{eq:arg}) to hold. In other words, the presence of zeroes in the coupling function is protected by inversion symmetry. In the particular case of real $V_j^B = V$ we obtain the nodes shown in Figs.\ \ref{fig:thetasq}(c) and \ref{fig:thetasq}(d).

A specific case in which the condition for inversion symmetry is not met is $V_j^A = 0$, $V_j^B = V \in \mathbb{R}$. In that case we obtain the coupling model for symmetric vacancies Eq.\ (\ref{eq:vVAC}), such that
\begin{equation}
    |\Theta_V^\pm(\vec{k})|^2 = \frac{V^2}{\sqrt{2}}|\Phi(\vec{k})|^2,
\end{equation}
has zeroes only at the $K$ and $K'$ points, protected only by $C_{3v}$ symmetry.

\bibliography{bibliography}

\end{document}